%% file: ms.tex
\documentclass[useAMS, usenatbib]{mnras}
\usepackage{color}
\usepackage{booktabs}   % Fancier tables
\usepackage{tabularx}
\usepackage{float}
\usepackage{totcount}
\usepackage{cuted}
\usepackage[dvipsnames, table]{xcolor}
\usepackage{graphicx}
\usepackage{multirow}
\usepackage[section]{placeins}
\usepackage{subfig}
\usepackage{multicol}
\usepackage{widetext}
\usepackage{upgreek}    % allow alternative greek letters like $\uptau$
\usepackage{enumitem}   % customize enumerate symbols
\setlist[enumerate]{leftmargin=*}   % Set left-margin of enumerate lists to match the edge
\setlist[itemize]{leftmargin=*}   % Set left-margin of enumerate lists to match the edge
\usepackage[T1]{fontenc}

\usepackage{lipsum} % generate dummy text
\usepackage{amsmath}
\usepackage{amssymb}

\graphicspath{ {figs/} }

\input{CustComs}

\input{JournalAbbrevs}

\usepackage{etoolbox}
\makeatletter
\patchcmd\@combinedblfloats{\box\@outputbox}{\unvbox\@outputbox}{}{\errmessage{\noexpand patch failed}}
\makeatother
\newcommand{\hvaramp}{\chi}
\newcommand{\per}{p}
\newcommand{\inc}{i}
\newcommand{\incmin}{i_\trt{min}}
\newcommand{\snr}{\mathrm{SNR}}

\newcommand{\fnu}{F_{\nu}}
\newcommand{\fnusens}{F_{\nu,\trt{sens}}}
\newcommand{\fnusenscrts}{F_{\nu,\trt{sens}}^\trt{CRTS}}
\newcommand{\fnusenslsst}{F_{\nu,\trt{sens}}^\trt{LSST}}
\newcommand{\dfnu}{\delta_{\tiny F}}
\newcommand{\dfnusens}{\delta_{{\tiny F}, \textrm{\tiny sens}}}
\newcommand{\dfnumin}{\delta_{{\tiny F}, \textrm{\tiny min}}}
\newcommand{\dfdop}{\delta_{\tiny F}^\trt{dop}}
\newcommand{\dfhyd}{\delta_{\tiny F}^\trt{hyd}}

\newcommand{\numsimbin}{N^\trt{sim}_\trt{bin}}
\newcommand{\numsimagn}{N^\trt{sim}_\trt{AGN}}

\newcommand{\numobsagn}{N^\trt{obs}_\trt{AGN}}

\newcommand{\numsimagncrts}{N^\trt{sim,CRTS}_\trt{AGN}}
\newcommand{\numobsagncrts}{N^\trt{obs,CRTS}_\trt{AGN}}

\newcommand{\numsimagnlsst}{N^\trt{sim,LSST}_\trt{AGN}}
\newcommand{\numobsagnlsst}{N^\trt{obs,LSST}_\trt{AGN}}

\newcommand{\fcomplsst}{f_\trt{complete}^\trt{LSST}}

\newcommand{\fracobsdop}{f_\trt{obs}^\trt{dop}}

\newtotcounter{citnum} % From the package documentation
\def\oldbibitem{} \let\oldbibitem=\bibitem
\def\bibitem{\stepcounter{citnum}\oldbibitem}

\title[MBH Binaries as Variable AGN]{Massive BH Binaries as Periodically-Variable AGN}
\author[L.Z.~Kelley et al.]{
    \hspace{-0.054in}Luke Zoltan Kelley$^{1,2}$\thanks{E-mail:lzkelley@northwestern.edu},
	Zolt\'an Haiman$^{3}$,
    Alberto Sesana$^{4}$,
	Lars Hernquist$^{1}$,
    \vspace{0.1in} \\
    $^{1}$ \begin{minipage}[t]{\linewidth} Harvard-Smithsonian Center for Astrophysics; Harvard University, Cambridge, MA 02138, USA \end{minipage} \\
    $^{2}$ \begin{minipage}[t]{\linewidth} Center for Interdisciplinary Exploration and Research in Astrophysics (CIERA); Northwestern University,
Evanston, IL 60208 \end{minipage} \\
    $^{3}$ \begin{minipage}[t]{\linewidth} Department of Astronomy, Columbia University, New York, NY 10027, USA \end{minipage} \\
    $^{4}$ \begin{minipage}[t]{\linewidth} School of Physics and Astronomy and Institute of Gravitational Wave Astronomy\\ \indent University of Birmingham, Edgbaston B15 2TT, UK\end{minipage}
}

\begin{document}

\pagerange{\pageref{firstpage}--\pageref{lastpage}} \pubyear{2018}

\maketitle
\label{firstpage}

\begin{abstract}
    Massive black-hole (MBH) binaries, which are expected to form following the merger of their parent galaxies, produce gravitational waves which will be detectable by Pulsar Timing Arrays at nanohertz frequencies (year periods).  While no confirmed, compact MBH binary systems have been seen in electromagnetic observations, a large number of candidates have recently been identified in optical surveys of AGN variability.  Using a combination of cosmological, hydrodynamic simulations; comprehensive, semi-analytic binary merger models; and analytic AGN spectra and variability prescriptions; we calculate the expected electromagnetic detection rates of MBH binaries as periodically variable AGN.  In particular, we consider two independent variability models: (i) Doppler boosting due to large orbital velocities, and (ii) hydrodynamic variability in which the fueling of MBH accretion disks is periodically modulated by the companion.  Our models predict that numerous MBH binaries should be present and distinguishable in the existing data.  In particular, our fiducial models produce an expectation value of $0.2$ (Doppler) and $5$ (hydrodynamic) binaries to be identifiable in CRTS, while $20$ and $100$ are expected after five years of LSST observations.  The brightness variations in most systems are too small to be distinguishable, but almost $1\%$ of AGN at redshifts $z \lesssim 0.6$ could be in massive binaries.  We analyze the predicted binary parameters of observable systems and their selection biases, and include an extensive discussion of our model parameters and uncertainties.
\end{abstract}

\begin{keywords}
quasars: supermassive black holes, galaxies: kinematics and dynamics
\end{keywords}

% Sec 1
\section{Introduction}
    \label{sec:intro}

    Active Galactic Nuclei (AGN) are known to often be triggered by interactions and mergers between their host galaxies \citep[e.g.][]{2000MNRAS.311..576K, comerford2015, Barrows201703, Goulding201706} which drive large amounts of gas towards the galaxy cores and Massive Black-Holes (MBHs) within \citep{barnes1992}.  Many examples of ``dual-AGN'', pairs of observably accreting MBHs in the same system, have been identified in radio, optical and X-ray surveys \citep[e.g.][]{rodriguez2006,Komossa2006,koss2012,comerford2012}.  After a galaxy merger, the two MBHs are expected to sink towards the center of the post-merger galaxy due to dynamical friction, which is very effective on $\sim 10^3$ pc scales \citep{Begelman1980, am12}.  Once the MBHs reach $\sim$ pc separations and smaller, and eventually become gravitationally bound as a MBH Binary (MBHB), the continued merging of the system depends sensitively on individual stellar scatterings extracting energy from the binary \citep[e.g.][]{Sesana200612, Merritt200705}.

    The effectiveness of stellar scattering in `hardening' MBH binaries remains unresolved, although studies are beginning to reach a consensus that the population of stars available for scattering (the `loss-cone') is efficiently refilled \citep[e.g.][]{sesana2015, vasiliev2015}.  Of particular interest is whether and which systems are able to reach the $\lesssim 10^{-3}$--$10^{-1}$ pc separations\footnote{for masses $\sim 10^6$ -- $10^9 \, \msol$.} at which point Gravitational Wave (GW) emission can drive the systems to coalesce within a Hubble time \citep{Begelman1980}.  While dual-AGN have been observed, there are no known AGN in confirmed gravitational bound binaries.  If MBHBs are able to reach periods of $\sim$ yr (frequencies $\sim$ nHz), their GW emission should be detectable by pulsar timing arrays \citep[PTAs;][]{hellings1983, foster1990}---the European \citep[EPTA,][]{desvignes2016}, NANOGrav \citep{arzoumanian201505}, Parkes \citep[PPTA,][]{reardon2016}, and the International PTA \citep[IPTA,][]{Verbiest201602}.  The most recent and comprehensive models for the cosmological population of merging MBHBs suggest that PTAs will plausibly make a detection within roughly a decade \citep[e.g.][]{taylor2015, Rosado1503, paper2}, and indeed, the most recent PTA upper-limits on GW signals---particularly on the presence of a power-law, Gravitational-Wave Background (GWB) of unresolved, cosmological sources---have already begun to inform the astrophysical models (\citealt{Simon201603}, \citealt{Taylor201612}; but also, \citealt{middleton201707}).

    MBH Binaries form on sub-parsec scales, which, even using VLBI, can only be spatially resolved at relatively low redshifts \citep[e.g.][]{DOrazio201712}.  Spectroscopic, and especially photometric methods, which don't require binaries to be spatially resolved, have recently put forward large numbers of binary candidates \citep{Eracleous201106, Tsalmantza201106, graham2015, charisi2016}.  On the theoretical side, few predictions have been made for the expected observability and detected rates of AGN in binary systems.  On kpc scales, \citet{2016MNRAS.458.1013S} study dual- and offset- AGN during the (relatively) early stages of galaxy merger, providing interesting results on the nature and properties of the MBH in these systems.  \citet{Volonteri2009} make rate predictions for tight MBHB, especially those with large orbital velocities that could be observable as dual broad-line AGN \citep[e.g.][]{Boroson2009}.  The authors predict an upper-limit to the detection rate of such systems between $0.6$ -- $1.0\E{-3}$ per unabsorbed AGN.

    The focus of this investigation are binaries and candidates identified by periodic variability in photometric surveys of AGN.  In particular, \citet{graham2015} find 111 candidates in $\sim 240{,}000$ AGN using the CRTS survey; \citet{charisi2016} find 33 in $\sim 35{,}000$ AGN using PTF; and \citet{Liu201609} initially identify 3 candidates in 670 AGN using PanSTARRS, however none are persistent in archival data.  These three detection rates are roughly consistent at $5\E{-4}$, $9\E{-4}$ and $\lesssim 4\E{-3} \, \invagn$.  It is worth noting that these detection rates are very similar to the limits from orbital-velocity selected populations in \citet{Volonteri2009}, which have similar binary parameters to the orbital-period selection for variability candidates.

    The connection between EM and GW observations of MBHBs has already begun to be leveraged using these photometric-variability candidates.  While none of the individual candidate systems can be excluded by PTA measurements, \citet{sesana201703} demonstrate that the \textit{population} of MBHBs that they imply leads to a GWB amplitude in tension with existing PTA upper-limits.  In \citet{sesana201703}, a phenomenological approach, with as few physical assumptions as possible, is used to connect EM and GW observations.  In this followup analysis, we rely instead on physically motivated, theoretical models to explore the repercussions on EM observations.  Specifically, we use binary populations based on the Illustris hydrodynamic, cosmological simulations \citep[e.g.][]{vogelsberger2014b, nelson2015} coupled with comprehensive semi-analytic merger models \citep{paper1, paper2} and synthetic AGN spectra to make predictions for the occurrence rates of periodically variable AGN.  In \secref{sec:meth} we summarize the binary population, the AGN spectra we use to illuminate them, and the models of variability we consider.  In \secref{sec:res} we present our results of expected detection rates and the parameters which determine binary observability.  Finally in \secref{sec:disc} we discuss the limitations of our study and discuss its implications for identifying and confirming MBH binaries through photometric variability studies.

% Sec 2
\section{Methods}
    \label{sec:meth}

    % Sec 2.1
    \subsection{MBH Binary Population and Evolution}
        \label{sec:meth_mbhb}

        Our MBHB populations are based on the MBHs and galaxies in the Illustris simulations.  Illustris is an $\left(108 \, \mathrm{Mpc}\right)^3$ volume of gas-cells and particles representing dark matter, stars, and MBHs which is evolved from the early universe to redshift zero \citep[e.g.][]{vogelsberger2014a, vogelsberger2014b, genel2014, torrey2014, rodriguez-gomez2015, nelson2015}.  The simulations include sub-grid models for star formation, stellar nucleosynthesis \& metal enrichment, and stellar \& AGN feedback.  MBH particles are initialized with a seed mass of $\sim 10^5 \, \msol$ in massive halo centers, after which they grow via accretion of local gas using a Bondi model.  Details of the BH prescription and resulting MBH and AGN populations are presented in \citet{sijacki2015}.  In the Illustris simulations, after or during a galaxy merger, once MBHs come within $\sim 10^2$ -- $10^3 \textrm{ pc}$ of one-another---roughly their gravitational smoothing length---they are manually merged and moved to the potential minimum of the halo.  To more carefully examine the MBHB merger process and dynamics, we `post-process' the MBH mergers using semi-analytic models.

        In this section we outline some of the key components of the merger models and the resulting merger dynamics that are described thoroughly in \citet{paper1, paper2}.  MBH-MBH ``merger-events" are identified in Illustris on $\sim \textrm{kpc}$ scales.  We then consider each of these events independently by extracting the MBH masses, and spherically-averaged galaxy density and velocity profiles for each constituent (dark matter, stars, gas) of the host.  These profiles are then used to calculate hardening rates of the semi-major axis ($da/dt$) based on prescriptions for dynamical friction \citep{chan42, bt87}, stellar `loss-cone' scattering \citep{magorrian1999}, viscous drag from a circumbinary disk \citep{hkm09, Tang201703}, and GW emission \citep{peters1963, Peters1964}.  Dynamical friction is required to harden the system on \mbox{$10$ -- $10^3 \textrm{ pc}$} scales, after which stellar scattering is typically dominant until the GW-dominated regime on \mbox{$\sim 10^{-2}$ -- $10^{-4} \textrm{ pc}$}.  In some systems, viscous drag can be the primary hardening mechanism near $\sim 10^{-2} \textrm{ pc}$.  Our population of binaries is roughly $10^4$ systems with total masses between $2\E{6} \, \msol$ and $2\E{10} \, \msol$, with a steeply declining mass-function.  The mass ratio of the systems is inversely correlated with total mass: systems with low total-masses can only have near-equal mass ratios due to the minimum MBH mass, and high total-mass systems are dominated by extreme mass-ratio mergers.

        Both stellar scattering and viscous drag remain highly uncertain processes.  The largest uncertainty affecting merger outcomes is likely the effectiveness of stellar scattering: in particular, how efficiently the stellar `loss-cone'---those stars  able to interact with the binary---are repopulated.  Typical coalescence lifetimes are gigayears.  Binaries which are both very massive $M \equiv M_1 + M_2 \gtrsim 10^9 \, \msol$, and near equal mass ratio $q \equiv M_2/M_1 \gtrsim 0.1$, are generally able to coalesce within a Hubble time.  Systems with both lower total masses ($M \lesssim 10^8 \, \msol$), and more extreme mass ratios ($q \lesssim 10^{-2}$) often stall at either kpc or pc separations\footnote{E.g.~for binaries with $M \lesssim 10^7 \, \msol$, $\sim 30\%$ of $q \gtrsim 0.3$ systems coalesce before redshift zero, and only $\sim 10\%$ of those with $q \lesssim 0.3$.  For the latter, low mass-ratio systems, this is unsurprising as dynamical friction is often ineffective at hardening to below $\sim$kpc separations \citep[e.g.][]{mcwilliams2014}.  In our models, we find that more comparable mass systems are more likely to stall at $\sim$pc scales, with stellar scattering becoming ineffective---likely due to less centrally-concentrated stellar mass distributions (see below).}.  The fate of the remaining, intermediate systems depends more sensitively on the assumed dynamical parameters (i.e.~the loss-cone refilling rate).  Note that this differs from some previous studies finding that more-massive systems merge \textit{less} effectively \citep[e.g.][]{yu2002, Cuadra200809, Dotti2015}\footnote{The discrepancy between Illustris-based models and those of some semi-analytic merger populations likely has to do with typical stellar densities in the cores of low mass galaxies.  In particular, the densities from Illustris may be systematically lower, but this has yet to be examined in detail.  The fraction of observable systems with low masses ($M \lesssim 10^7 \, \msol$) drops rapidly (e.g.~\figref{fig:10_rate-comp}), implying that this difference between models has little effect on our results.}.  The systems which reach $\sim \yr$ periods are thus somewhat biased against low total-masses, and extreme mass-ratios.

        Predictions for the GWB and its prospects for detection by PTAs are presented in \citet{paper2}, along with a description of our formalism for eccentric binary evolution.  Most models predict GWB amplitudes at periods of $1$ yr, $\ayr \approx 0.5$ -- $0.7 \E{-15}$, roughly a factor of 2 below current sensitivities, and detectable within about a decade.  Predictions for GW signals from individually resolvable `single-sources' are presented in \citet{paper3}, and are comparable in detectability to the GWB.  The results we present in this paper are relatively insensitive to variations in binary evolution parameters, compared to those of the electromagnetic and observational models we describe below.  For reference, the evolutionary model used here assumes an always full loss-cone and initial binary eccentricities of $\eccinit = 0.5$.

    % Sec 2.2
    \subsection{MBH Accretion and AGN Spectra}
        \label{sec:meth_spectra}

        % Fig 1 - Accretion ratio / accretion partition function
        \begin{figure}
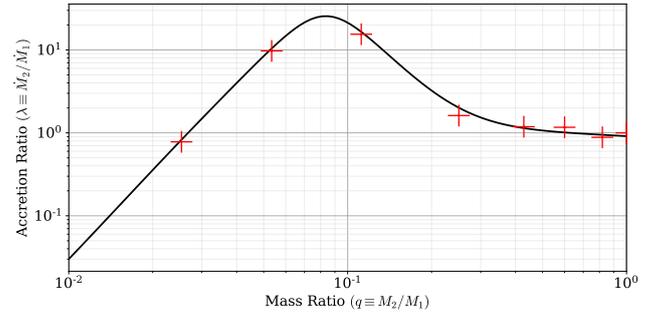

            \centering
            \includegraphics[width=\columnwidth]{{{figs/1_acc-ratio}}}
            \caption{Accretion ratio data points are from hydrodynamic simulations of MBHB in circumbinary accretion disks by \citet{farris201310}.  The line is fit with the function and parameters that are shown in \refeq{eq:acc_rat}.}
            \label{fig:1_acc_ratio}
        \end{figure}

        Our merger models follow the constituent MBHs of a given binary for long after it has ``merged" in Illustris.  After the ``merger'', Illustris records the accretion rate of ambient gas onto the single, remnant MBH.  We use this accretion rate as a measure of the fueling to the binary system as a whole, feeding the circumbinary disk: $\dot{M} = \dot{M}_1 + \dot{M}_2$.  In our post-processing models, however, the two MBH are \textit{un}merged, leaving an ambiguity in the feeding rate to each individual component.  To resolve this, we use the results from the detailed circumbinary-disk simulations in \citet{farris201310}, which give the ratio of accretion rates for a variety of binary mass-ratios: $\mdotrat = \mdotrat(q) \equiv \dot{M}_2 / \dot{M}_1$.  The simulation data-points are plotted in \figref{fig:1_acc_ratio}, along with a fit described by the function,
        \begin{align}
            \label{eq:acc_rat}
            \begin{split}
            \mdotrat = q^{a_1} e^{-a_2/q} + \frac{a_3}{\left(a_4 q\right)^{a_5} + \left( a_4 q \right)^{-a_5}} \\
            a_1 = -0.25,  \,\,  a_2 = 0.1,  \,\,  a_3 = 50,  \,\,  a_4 = 12,  \,\,  a_5 = 3.5.
            \end{split}
        \end{align}
        We assume that the system is Eddington limited on large scales, i.e.~\mbox{$\dot{M} \leq \mdotedd \equiv \dot{M}_{\textrm{Edd},1} + \dot{M}_{\textrm{Edd},2}$}, where $\mdotedd = 1.4\E{18} \textrm{ g s}^{-1} \, \scale{M}{\msol} \, \scale[-1]{\radeff}{0.1}$, and $\radeff$ is the radiative efficiency which we take as $0.1$.  We let each MBH individually exceed Eddington\footnote{The alternative is explored in \secref{sec:app_pars}.} \citep[e.g.][]{Jiang2014} which can occur for the secondary when $\mdotrat > 1.0$, corresponding to $q \gtrsim 0.03$.  The secondary accretion rate is maximized at $\lambda_\textrm{max} = \lambda(q \approx 0.08) \approx 25$.

        % Fig 2 - AGN Spectra
        \begin{figure}
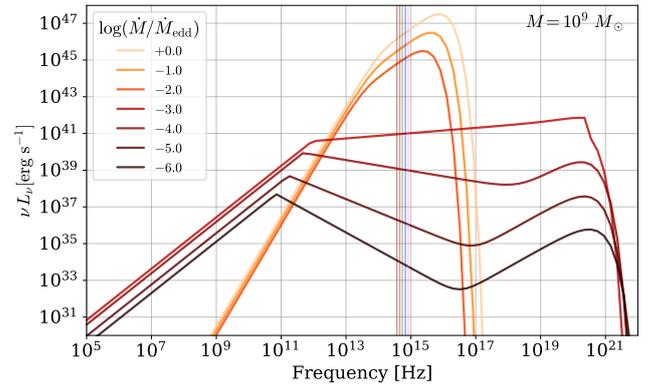

            \centering
            \includegraphics[width=\columnwidth]{{{figs/2_agn-spectra}}}
            \caption{AGN spectra for an MBH of mass $M = 10^9 \, \msol$, and a variety of accretion rates.  For Eddington ratios $\fedd \equiv \mdot / \mdotedd < 10^{-2}$, we use the ADAF emission model from \citet{Mahadevan1997}, while for larger accretion rates we use a thermal, Shakura-Sunyaev spectrum.  For reference, the vertical colored lines are the optical bands: $[i, r, g, v, b, u]$.}
            \label{fig:2_agn_spec}
        \end{figure}

\begin{figure}
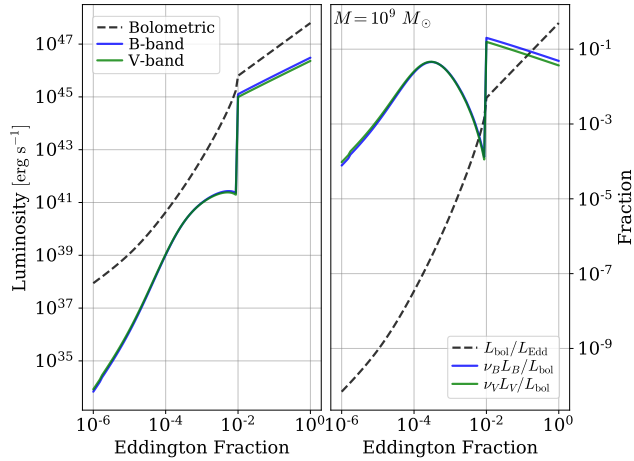

            \centering
            \includegraphics[width=\columnwidth]{{{figs/3_lum-bol-agn_band-bv_cols}}}
            \caption{Luminosity and radiative efficiency versus Eddington ratio.  The left panel shows bolometric (dashed), B-band and V-band luminosities (solid lines), calculated for a $M=10^9 \, \msol$ MBH, against a variety of accretion rates.  The right panel gives the overall radiative efficiency \mbox{$\radeff = L_\textrm{bol} / \ledd$}, as well as the fraction of energy emitted in the B- and V- bands (i.e.~the inverse of the bolometric corrections).}
            \label{fig:3_agn_lbol}
        \end{figure}

        The parameters for MBH evolution in Illustris are calibrated to match the observed M--$\sigma$ relation, and the AGN (bolometric) luminosity function based on a constant radiative efficiency of $0.05$ \citep[see,][]{sijacki2015}.  For our analysis, we calculate full spectra for each MBH based on its mass and Eddington ratio, $\fedd \equiv \mdot / \mdotedd$, obtained from the accretion rates described above.  For $\fedd \geq 10^{-2}$, we assume the accretion flow is radiatively efficient and use a \citet{shakura1973} `thin'-disk solution, which assumes emission is purely thermal from each annulus of the disk.  For $\fedd < 10^{-2}$ we assume radiatively inefficient accretion in the form of an ADAF \citep{Narayan1995b}, and use the emission model from \citet{Mahadevan1997}.  The ADAF model includes self-absorbed synchrotron emission, bremsstrahlung, and inverse-Compton of synchrotron photons.  We thus calculate AGN spectra as,
            \begin{align}
                \label{eq:spec}
                F_\nu   & = F_\nu^\trt{thin}(M, \fedd) \hspace{0.01in} & \fedd \geq 10^{-2}, \\
                        & = F_\nu^\trt{ADAF}(M, \fedd) \hspace{0.01in} & \fedd < 10^{-2}.
            \end{align}
            Spectra for a variety of accretion rates onto an $M = 10^9 \, \msol$ BH are shown in \figref{fig:2_agn_spec}.  The bolometric luminosity and the luminosities in the B- and V- band are shown in \figref{fig:3_agn_lbol}, along with the effective radiative-efficiency ($L_\textrm{bol} / L_\mathrm{Edd}$) and luminosity fractions (the inverse of the bolometric corrections).  The change in spectral shape and optical luminosity at the transition of $\fedd = 10^{-2}$, corresponding to the change from thick to thin accretion flows, is clear in both \figref{fig:2_agn_spec}~\&~\figref{fig:3_agn_lbol}.  While this transition is likely artificially abrupt, it is consistent with transitions in the state of X-ray binaries, associated with the same changes in accretion regime \citep[e.g.][]{Esin1997}.  Regardless, the specific location and sharpness of this transition has little effect on our results as observable systems are predominantly at $\fedd \gtrsim 0.1$.

        % Fig 3 - AGN Bolometric correction

\begin{figure}
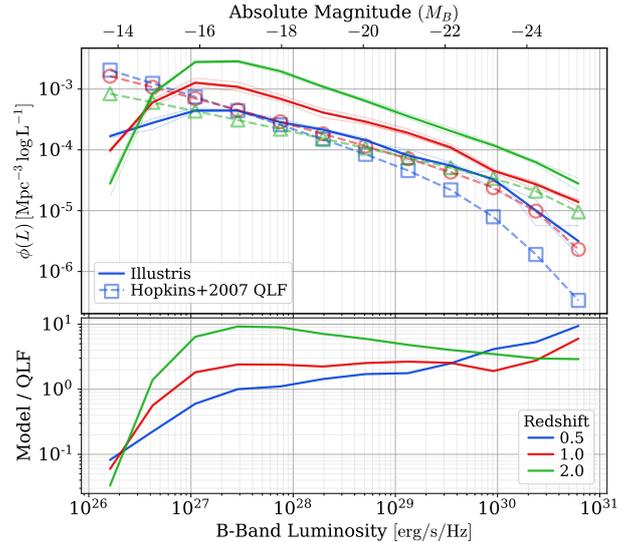

            \centering
            \includegraphics[width=\columnwidth]{{{figs/4_agn-func_lum}}}
            \caption{AGN luminosity functions constructed from Illustris MBH (solid lines) and the observationally-derived quasar luminosity function (QLF) from \citet[][points / dashed-lines]{Hopkins2007}.  The top panel shows the luminosity functions at three different redshifts, \mbox{$z = 0.5$, $1.0$ \& $2.0$}, and the bottom panel shows the ratio of QLFs between Illustris and observations.  While our models are consistent with observations to an order of magnitude, our MBH populations and synthetic spectra noticeably over-predict the luminosity function.  Note that this comparison is shown for the B-band, while most of our analysis focuses on the V-band.}
            \label{fig:4_agn_lumfunc}
        \end{figure}

        The luminosity function of Illustris AGN, using our spectral models, is compared to the observationally determined quasar luminosity function (QLF) from \citet{Hopkins2007} in \figref{fig:4_agn_lumfunc}.  Comparing observed and simulated AGN observations is non-trivial.  On the observational side there are extinction, bolometric corrections, K-corrections, and selection biases; while on the simulation side, we are using disk-integrated quantities based on semi-analytic models instead of either radiative transfer calculations or full disk-simulations.  None the less, our models agree with observations to well within an order of magnitude, although the Illustris AGN tend to be systematically over-luminous compared to the observed QLF.  Throughout our analysis we present our results primarily in terms of observable binary detections normalized to the predicted number of observable AGN.  This is to reduce errors from our reproduction of the luminosity function.  In \secref{sec:app_pars} we also present results for a model in which all accretion rates have been lowered by a factor of three, which does produce a better match to the observed QLF.

        % Fig 4 - AGN Luminosity Function

        The luminosity functions from \citet{Hopkins2007}, shown in \figref{fig:4_agn_lumfunc}, correct for obscuration.  In our analysis, we use the same model which assumes that a luminosity-dependent fraction of systems are observable,
            \begin{equation}
                \begin{split}
                f(L) = \min \left[ 1, \, f_{46} \left( \frac{L}{10^{46} \textrm{ erg s}^{-1}}\right)^\beta\right], \\
                \textrm{\textit{B-band}:} \,\,\,\, f_{46} = 0.260, \,\, \beta = 0.082,
                \end{split}
            \end{equation}
            where $L$ is the bolometric luminosity, and the fit values are for the B-band.  We convert between bolometric and spectral luminosity using a luminosity-dependent bolometric correction \citep{Hopkins2007},
            \begin{equation}
                \begin{split}
                \frac{L}{L_i} = c_1 \left( \frac{L}{L_\odot}\right)^{k_1} + c_2 \left( \frac{L}{L_\odot}\right)^{k_2}, \\
                \textrm{\textit{B-band}:} \,\,\,\, c_1 = 7.40, \,\, c_2 = 10.66, \,\, k_1 = -0.37, \,\, k_2 = -0.014.
                \end{split}
            \end{equation}
        This bolometric correction, which would differ in general from that in our spectral models, is used here for the sole purpose of computing the obscuration fraction in a way consistent with \citet{Hopkins2007}. It is not used elsewhere in our analysis.

\begin{figure*}
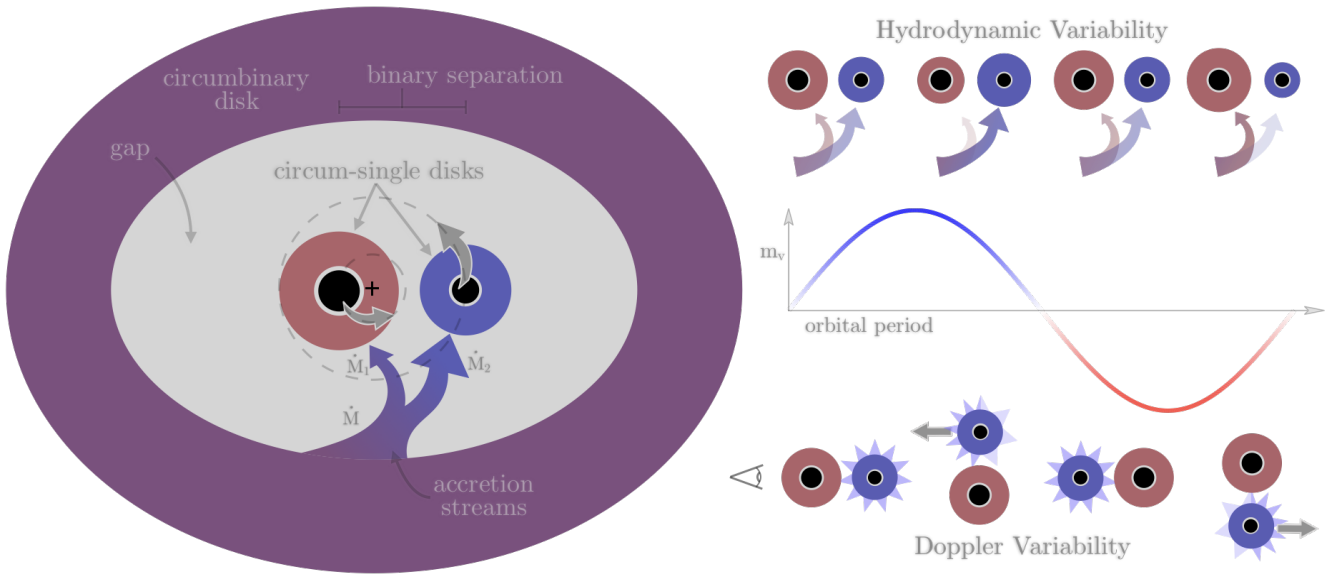

            \centering
            \includegraphics[width=\textwidth]{{{figs/0_schematic}}}
            \caption{Schematic representation of the binary, disk, and accretion geometries assumed in our models; informed from the results of hydrodynamic simulations \citep[e.g.][]{farris201310}.  \textit{Left:} between the binary and the circumbinary disk is a `gap' with a radius roughly twice the binary separation.  Around each MBH is a `circum-single' disk, fed by time-variable accretion streams extending from the circumbinary disk.  Because the secondary MBH is farther from the center-of-mass, and closer to the circumbinary disk, it tends to receive a disproportionate share of the accretion rate.  \textit{Right:} the hydrodynamic and Doppler mechanisms for producing photometric variability are depicted on the top and bottom respectively.  The circumbinary disk orbits at longer periods than the circum-single disks that it feeds, causing periodic variations in accretion rate, and thus luminosity.  For observers oriented near the orbital plane, Doppler boosting of the faster moving, and typically more luminous, secondary MBH can also produce brightness variations.}
            \label{fig:0_schematic}
        \end{figure*}

        One additional adjustment is required for spectra from disks in binary systems: the presence of a companion leads to truncation of each ``circum-single'' disk at a radius comparable to the Hill-radius.  Specifically we set the outer-edge of each disk to,
        \begin{align}
            \label{eq:disk-truncate}
            r_\trt{max} & = \frac{a}{2} \, \scale[1/3]{M_i}{M}, \\
                        & \sim 74 \, \rs \scale[2/3]{\per}{5 \, \yr} \, \scale[-2/3]{M}{10^9 \, \msol}
        \end{align}
        where $M_i$ is the mass of the primary or secondary, $a$ is the semi-major axis, $P$ is the orbital period and $\rs$ is the Schwarzschild radius of the MBH in question.  A typical geometry for the binary is shown schematically in \figref{fig:0_schematic}, showing the two circum-single disks around each MBH, within a larger circumbinary disk.  While the bright, optical emission in AGN tends to come from relatively small radii, disk truncation can be important for especially massive BHs and those in short-period binaries.  For example, the optical luminosity of a $10^9 \, \msol$ MBH, in a $5 \, \yr$ period binary, can be decreased by $\sim 10$ -- $20\%$.  When calculating the total luminosities of binaries, we also include contributions from the circumbinary portion of the accretion disk, where, fiducially, we assume that the inner-edge occurs at twice the binary separation, and the inner-edge of each circum-single disk is located at $3 \, \rs$ (i.e.~the inner-most stable circular orbit for a non-spinning BH).

    % Sec 2.3
    \subsection{Populations and Observations}
        \label{sec:meth_obs}

        We wish to calculate the number of binaries observable at a given log-interval of period, and a particular interval of redshift, $d^2N / dz \, d\logten \per$.  By using the number density in a comoving volume, $n \equiv dN / dV_c$, we can write,
            \begin{equation}
                \frac{d^2 N}{dz \, d\logten \per} = \fracobs \, \frac{\per}{\ln 10}  \, \frac{dn}{d\per} \frac{dV_c}{dz}.
            \end{equation}
        We have introduced the factor $\fracobs$, the fraction of systems that are observable at a given redshift, which is generally a function of any binary parameter (e.g.~mass, orbital period, inclination, etc).

        We consider particular periods of interest $\per_j$, and for each simulated binary $i$ in our population, we find the redshift at which it reaches those periods: $z_{ij}$.  To calculate the number or number density of sources, we consider discrete bins in redshift, $\Delta z'_k \in [z'_k, z'_{k+1})$, and identify all binaries reaching the period of interest in that bin.   A given binary will be counted at all periods that it reaches before redshift zero.  In this way, we effectively treat each moment in time, for each binary, as a separate data sample---i.e.~each represents an independent population of astrophysical binaries (at a slightly different redshift).

        Because we are sampling in orbital period, instead of evenly in time, we must explicitly account for the time evolution of binaries by considering the fraction of time binaries spend emitting at each period.  The temporal evolution of binaries can be written as,
            \begin{equation}
                \frac{dn}{d\per} = \frac{dn}{dt} \, \frac{dt}{df} \frac{df}{d\per}.
            \end{equation}
        Using Kepler's law along with the hardening time, $\thard \equiv a / (da/dt)$, we can write,
        	\begin{equation}
        	       \frac{dt}{df} = \frac{2}{3} \thard \, \per.
        	\end{equation}
        The number of binaries at period $\per = \per_j$, in redshift bin $k$, is then,
            \begin{align}
                \begin{split}
                \frac{d N_{jk}}{d\logten \per_j} & = \sum_i \frac{\fracobsi}{\ln 10} \cdot \delta(z'_k \leq z_{ij} < z'_{k+1})  \cdot \mathcal{T}_{ijk} \cdot \mathcal{V}_{k}, \\
                \mathcal{T}_{ijk} & \equiv \min\left(\frac{2}{3} \frac{\thardij}{\Delta t_k}\, , \, 1\right) ,\\
                \mathcal{V}_{k} & \equiv \frac{1}{\volill} \frac{dV_c}{dz} \Delta z'_k.
                \end{split}
            \end{align}
        The differential, comoving volume of the universe as a function of redshift, $dV_c / dz$, is given in \citet[][Eq.~28]{hogg1999}.

        The fraction of systems observable at a given redshift and period, $\fracobs$, depends on the emission and variability model.  We require that the flux from the source is above the flux threshold, $\fnu \geq \fnusens$, and that the variability, $\dfnu \equiv \Delta \fnu / \fnu$, is above the variability threshold, $\dfnu \geq \dfnusens$, i.e.,
            \begin{equation}
                \fracobs = \Theta(\fnu \geq \fnusens) \cdot \Theta(\dfnu \geq \dfnusens).
            \end{equation}
        Here $\Theta$ is the Heaviside function: unity when the argument is true, and zero otherwise.  The variability threshold depends on the SNR and a minimum variability floor, $\dfnumin$.  Based on the smallest variability amplitudes seen in \citet{graham2015}, we use a fiducial value of $\dfnumin = 0.05$.  We calculate the variability threshold as,
            \begin{equation}
                \dfnusens = \snr^{-1} + \dfnumin,
            \end{equation}
        which is consistent with the results of variability studies in HST by \citet[][see their Fig.~3]{Sarajedini200308} and in PanSTARRS by \citet[][see their Fig.~4]{Liu201609}.  In both cases, the authors find minimum detectable variabilities of $\sim 1$ -- $2\%$, and then select systems using a cut which is some factor larger, at $\sim 5\%$.  Similarly, we calculate the SNR by assuming the flux-sensitivity threshold is a factor of five above the noise, i.e.~\mbox{$\snr = 5 \fnu / \fnusens$}.

        Our results are calculated for a range of detector sensitivities which encompass current and near-future instruments.  For convenience, we frequently present our results in terms of two characteristic values based on CRTS and LSST sensitivities.  In particular, we use V-band sensitivities of \mbox{$\fnusenscrts = 4\E{-28} \sfluxunits$} ($m_\trt{V} \approx 19.5$) and \mbox{$\fnusenslsst = 3\E{-30} \sfluxunits$} ($m_\trt{V} \approx 24.9$) for CRTS and LSST respectively\footnote{The CRTS value we get from the cutoff in the flux distribution of candidates from \citet{graham2015}, while the LSST value is from \citet{LSST2008}.  Our detection rates are not strongly dependent on the particular flux-threshold, as discussed in \secref{sec:app_pars}.}.  CRTS is considered in particular because of their large sample of binary candidates, but we consider this to be representative of current survey capabilities in general.

        The overall number-density of sources (i.e.~in units of $\textrm{Mpc}^{-3}$) is the metric most directly extracted from our models.  To better compare with observations, and to reduce the impact of systematic uncertainties in our luminosity functions, we focus on the number of simulated, observable binaries ($\numsimbin$) as a fraction of the expected number of all observable AGN ($\numsimagn$), which we report in units of $\invagn$.  To compute $\numsimagn$, we calculate the observability of all Illustris MBHs, using the same spectral models described in \secref{sec:meth_mbhb}, out to a redshift $\zmax = 4.0$.  For the aforementioned sensitivities, our fiducial simulations predict $10^6$ all-sky AGN to be observable by CRTS, and $4\E{7}$ by LSST.  CRTS actually observes $\approx 3\E{5}$ spectroscopically-confirmed AGN \citet{graham2015}.  The spectroscopic data comes primarily from SDSS, which covers only about one-quarter of the sky, suggesting an all-sky number a little above $10^6$, and consistent with our calculation\footnote{We emphasize, however, that this degree of consistency is largely fortuitous, and even somewhat surprising as Illustris tends to over estimate the AGN luminosity function (e.g.~\figref{fig:4_agn_lumfunc}).}.

        For a given sensitivity, our simulations predict an expected number of detected AGN and periodically-variable binaries.  A more robust prediction for a given survey can be calculated from the binary detection rate per AGN, along with the actual number of monitored AGN in the survey.  We refer to this prediction as a ``rescaled'' number of expected detections,
        \begin{equation}
        	\label{eq:rescale}
        	N^\trt{rescaled}_\trt{bin} = \numsimbin \, \frac{\numobsagn}{\numsimagn}.
        \end{equation}
        When making predictions for LSST, we assume a completeness $\fcomplsst = 2$, relative to CRTS\footnote{CRTS binary candidates are identified from SDSS spectroscopically confirmed AGN, which have a high completeness over roughly 1/4 of the sky \citep[e.g.][]{2009ApJS..180...67R}.  Studies suggest that through a combination of photometric and variability selection, LSST can achieve $>90\%$ completeness over its entire field of view \citep[e.g.][]{2010ApJ...714.1194S, 2011AJ....141...93B, 2011ApJ...728...26M}---roughly half of the sky, or twice that of SDSS (and thus the CRTS sample).}, i.e.,
    	\begin{equation}
    		\numobsagnlsst = \numsimagnlsst \frac{\numobsagncrts}{\numsimagncrts} \fcomplsst.
    	\end{equation}

        % Fig 0 - Accretion ratio / accretion partition function

    % Sec 2.4
    \subsection{Models of Variability}
        \label{sec:meth_var}

        The luminosity of an object in a binary system will not necessarily vary on the orbital period or at all.  The premise of photometric identification of MBH binaries is that the binary period is somehow imprinted into variations of the observed luminosity.  In the particularly convincing example of PG 1302-102 \citep{Graham201501}, sinusoidal variations in the light-curve can be well explained by Doppler-boosting from a mildly relativistic orbital velocity (\citealt{D'Orazio201509}; but see also \citealt{Liu201803}).  Additionally, purely hydrodynamic modulations to accretion rates have been observed in simulations \citep[e.g.][]{farris201310}.  Here we describe models for both types of variability mechanisms.  Throughout our analysis, each variability mechanism is considered entirely independently, i.e.~we do not consider systems which may be both hydrodynamically \textit{and} Doppler variable.

        The physical scenario is depicted schematically in \figref{fig:0_schematic}, assuming thin-disk geometries and a mass-ratio $q \gtrsim 10^{-2}$ (see \secref{sec:meth_var_hyd} below).  On the left, the two MBH, each with a circum-single disk, are shown within a cavity (`gap') of material evacuated by the binary orbit which separates them from the circumbinary disk.  The radii of the circum-single disks are determined by the Hill radius of each MBH (see \refeq{eq:disk-truncate}).  Despite the presence of the gap, the circumbinary disk continues to transport angular momentum outwards, requiring material to be accreted inwards.  Material from the disk overflows across the gap as accretion streams onto each circum-single disk.  The orbital period at the inner edge of the circumbinary disk is longer than that of the binary, causing periodic variations in the accretion rate onto each circum-single disk (\figref{fig:0_schematic}, upper-right).  The secondary MBH is farther from the center of mass and closer to the circumbinary disk edge, which leads to it receiving a disproportionate fraction of the accreting material (as described by \refeq{eq:acc_rat} and shown in \figref{fig:1_acc_ratio}).  Each circum-single disk, in addition to the circumbinary disk (although typically to a lesser extent), will produce AGN-like emission.

         % Sec 2.4.1
        \subsubsection{Doppler Variability}
            \label{sec:meth_var_dop}

            Any component of the binary orbital-velocity along the observer's line-of-sight will lead to both a relativistic boost and a Doppler shift in the observed spectrum of each circum-single disk.  The boost is calculated using the Doppler factor, \mbox{$D \equiv \left[ \gamma \left(1 - v_{||}/c\right)\right]^{-1}$}, where the Lorentz factor \mbox{$\gamma \equiv \left(1 - v^2/c^2\right)^{-1/2}$}.  The line-of-sight velocity, \mbox{$v_{||} \equiv v \sin(\inc)$}, depends on the binary inclination $\inc$, which we define as zero for face-on systems.  The observed flux is calculated as \citep{D'Orazio201509},
            % Eq.14
            \begin{equation}
                \label{eq:dop_1}
                F_{\nu} = D^3 F'_{\nu'},
            \end{equation}
            where the observed frequency $\nu$ is related to the rest-frame frequency as, $\nu = D \, \nu'$.  Assuming a power-law of index $\alpha_\nu$ for the section of the spectrum being observed, the Doppler variation in flux from the source will be \citep{Charisi2018},
            % Eq.15
            \begin{equation}
                \label{eq:dop_2}
                \frac{\Delta F_\nu^d}{F_\nu} = \left(3 - \alpha_\nu\right) \left|\frac{v}{c}\right| \sin \inc,
            \end{equation}
            where $v$ is the orbital velocity and $c$ is the speed of light.  The sensitivity of Doppler boosting to frequency and thus spectral shape offers a powerful method of testing it as a variability mechanism.  \citet{D'Orazio201509} have shown that, in both the optical and ultraviolet, this model explains the periodic variations observed in PG 1302-102.

            In full generality, an AGN spectra may not be a power-law at the frequency of interest, so we construct a full spectrum for each MBH in our simulations and numerically calculate the change in flux using \refeq{eq:dop_1}.  Additionally, the Doppler-boosting of each MBH in a binary is necessarily $\pi$ out of phase, thus we determine the overall system variation as,
            \begin{equation}
                \label{eq:var_doppler}
                \dfdop \equiv \frac{\Delta F_{\nu,1}^d - \Delta F_{\nu,2}^d}{F_{\nu,1} + F_{\nu,2}}.
            \end{equation}
            To handle the inclination dependence, for each simulated binary we calculate the SNR based on the un-boosted flux, $\fnu$, and determine the minimum observable inclination $\incmin$, such that the variability is observable.  The fraction of solid-angles at which the system is observable, which, for randomly oriented inclinations is $\cos(\incmin)$, then contributes linearly to the observability fraction, i.e.,
            \begin{equation}
                \fracobsdop = \delta(\fnu \geq \fnusens) \cdot \delta(\dfnu \geq \dfnusens) \cdot \cos(\incmin).
            \end{equation}

        % Sec 2.4.2
        \subsubsection{Hydrodynamic Variability}
            \label{sec:meth_var_hyd}

            Periodic variations in accretion rates are frequently observed in hydrodynamic simulations of circumbinary disks \citep[e.g.][]{1994ApJ...421..651A, 1996ApJ...467L..77A, Hayasaki200609, Roedig201202, 2013MNRAS.436.2997D, farris201310, Munoz2016}.  While significant uncertainties remain in understanding these accretion flows, the general pattern emerging is that three distinct mass-ratio regimes exist.  For extreme mass-ratios, $q \lesssim \qmin$, where $\qmin \sim 10^{-2}$, the secondary is a minor perturbation to the circumbinary disk, and the accretion flow remains steady.  At intermediate mass ratios, $\qmin \lesssim q \lesssim \qcrit$, where $\qcrit \approx 1/3$, a gap is opened by the secondary and the accretion rate onto it varies by a factor of order unity, on the binary orbital period \citep{D'Orazio2016}.

            For near-equal mass ratio systems ($q \gtrsim \qcrit$), a highly distorted cavity is evacuated around the binary, out to roughly twice the binary separation.  At the outer edge of the cavity, a significant over density of material develops \citep[see also:][]{2015ApJ...807..131S, 2012ApJ...755...51N}.  The Keplerian orbital period of that over density sets the variation timescale as $5$ -- $6$ times the binary period.  The binaries we are considering (i.e.~$M > 10^6 \, \msol$, $\torb \sim $ yr) are almost always in the GW-dominated regime in which the hardening timescale---the duration a given binary spends at that separation---decreases rapidly with decreasing orbital period.  Thus, if a given variational timescale is probing binaries at shorter periods, the number of observable systems decreases \citep{2015MNRAS.452.2540D}.

            We assume that hydrodynamic variability takes place for binaries in the thin-disk state, i.e.~$\fedd \geq 10^{-2}$.  While fluctuations in the accretion rate are also likely to occur for ADAF disks, the simulations exploring this phenomenon \citep[e.g.][]{D'Orazio2016} are primarily applicable to thin disks.  Modeling only variations in high-accretion rate systems has negligible effects on our overall results, as the low accretion rate systems are much harder to observe.  Based on \citet{farris201310}, we assume for our fiducial models that all binaries with mass ratios above $\qmin = 0.05$ exhibit hydrodynamic variability, and those above $\qcrit = 1/3$ are observable at $\tvar = 5 \, \torb$.  The accretion-rate variations in simulations predominantly effect the secondary MBH \citep[e.g.][]{farris201310}, so we model the overall hydrodynamic variations as,
            \begin{equation}
                \label{eq:var_hydro}
                \dfhyd \equiv \frac{\Delta F_{\nu,2}^h}{F_{\nu,1} + F_{\nu,2}} = \frac{ F_{\nu,2}(M_2, \hvaramp \feddtwo) - F_{\nu,2}(M_2, \feddtwo)}{F_{\nu,1} + F_{\nu,2}},
            \end{equation}
            where we take the effective enhancement to the accretion rate as $\hvaramp = 1.5$.  In \secref{sec:app_pars} we explore alternative values of $\hvaramp$, and the importance of the $\tvar$ assumption.

% Sec 3
\section{Results}
    \label{sec:res}

    % Fig 5 - Variability-Grid example
    \begin{figure*}
        \centering
        \includegraphics[width=2\columnwidth]{{{figs/5_var-grid_fe-1.00_per3.14_band-v_trunc-True_circum-2.0_var-adaf-False_annot-False}}}
        \caption{\textbf{Variability amplitudes in total-mass--mass-ratio space for Doppler and hydrodynamic variability.}  In this example, we use a fixed orbital period of \mbox{$\torb = 3.14 \, \yr$}, and a system accretion-rate of \mbox{$\feddsys = 10^{-1}$}.  The accretion rate onto the secondary is shown in the green color-bar to the right.  The hashed regions correspond to luminosities below the detection threshold for LSST (larger circles) and CRTS (smaller dots) at a uniform redshift \mbox{$z = 1$}.  Note that fiducially, based on the results of \citet{farris201310}, only systems with $q > \qmin = 0.05$ (red, dashed line) are considered as variables, although here we show variability at lower mass-ratios.  The contours correspond to the indicated variability amplitudes.}
        \label{fig:5_var-dop_grid}
    \end{figure*}

    Flux variability amplitudes for a grid of mass-ratio and total-mass are shown in \figref{fig:5_var-dop_grid} for both Doppler (left) and hydrodynamic variability (right).  Here we use fixed values of orbital period: \mbox{$\torb = 3.14 \, \yr$}, the system accretion-rate: \mbox{$\feddsys = 10^{-1}$}, and redshift: \mbox{$z = 1$}.  The accretion rate onto the secondary MBH, determined by the accretion partition function (\refeq{eq:acc_rat}), is illustrated by the green color-bar.  For both variability mechanisms, the secondary's variations tend to dominate across the parameter space, although for shorter periods and very high total-mass systems, the primary can occasionally dominate.

    Figure~\ref{fig:5_var-dop_grid} shows a sharp discontinuity in variability amplitudes at $q \approx 2.5\E{-3}$, corresponding to the point at which the secondary's accretion rate drops below $\fedd = 10^{-2}$, and its disk transitions to the ADAF state.  Note that the location at which this transition occurs is sensitive to the two simulated data points constraining the fitting function in this $q \ll 1$ domain (see \figref{fig:1_acc_ratio}).  In the case of hydrodynamic variability, recall that our model considers only variations occurring in the thin-disk state, and additionally, only systems with $q \ge \qmin = 0.05$ are considered as variable in our fiducial configuration.  Systems near this transitionary mass ratio will alternate between thin and ADAF disks, which produces very high variability amplitudes\footnote{This is the case even allowing for variability in the ADAF state (not shown).}.  Doppler variability shows a similar, although more mild, discontinuity at the same transition mass-ratio, due entirely to the drop in optical luminosity of the secondary as it transitions to the ADAF state.

    Doppler variability amplitudes are significantly larger for higher total-mass systems which have larger orbital velocities.  At the largest total masses, however, truncation of the secondary's accretion disk becomes significant, and overall variability amplitudes again decline.  Besides the mass-ratio and accretion-rate cutoffs imposed in our model, the hydrodynamic variability amplitudes have no explicit dependence on total mass or mass ratio.  The trends seen in the right panel of \figref{fig:5_var-dop_grid} are instead due to the relative brightness of the secondary compared to that of the primary and circumbinary disk.  At large total masses, the secondary's disk is significantly truncated, which decreases its luminosity and thus the overall variability amplitude.  When most of the emitting region of the secondary's accretion disk is preserved (i.e.~at lower total masses), the relative accretion rate onto the secondary determines its brightness and the variability amplitude.

    The total luminosity of the system is tied closely to its total mass.  Near $q \approx 0.1$, the accretion rate onto the secondary is enhanced by more than the inverse mass-ratio, leading to the secondary outshining the primary.  Near this mass-ratio sweet spot, systems remain observable even at lower total-masses.  The hatched regions of \figref{fig:5_var-dop_grid} show systems with V-band spectral fluxes below the LSST (circles) and CRTS (dots) thresholds for systems at a fixed luminosity distance of \mbox{$\dl(z=1) \approx 6.5 \textrm{ Gpc}$}.  At this distance, systems with $\fedd = 10^{-1}$ can be seen down to $M \approx 2\E{8} \, \msol$ and $M \approx 3\E{6} \, \msol$ for CRTS and LSST respectively.

    % 3.1
    \subsection{Event Rates}

        % Fig 6 - Number-Density vs. Redshift, by AGN, all-bins, flux-bins, var-bins
        \begin{figure*}
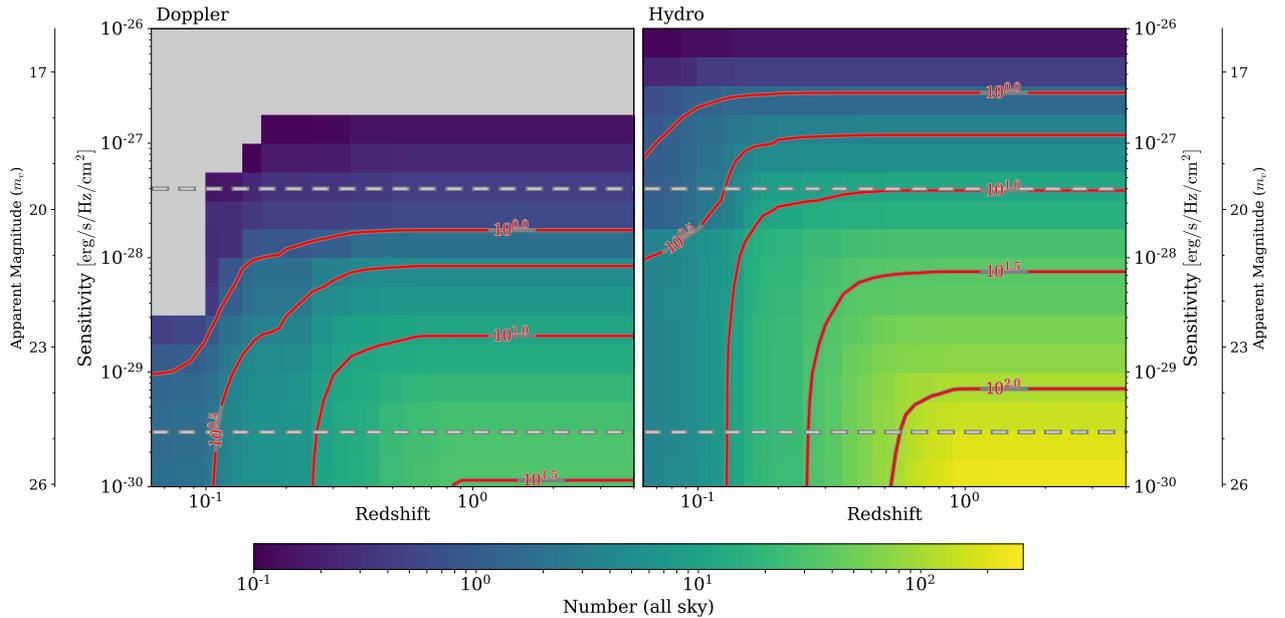

            \includegraphics[width=2\columnwidth]{{{figs/6_num-variables_vs_redshift-sensitivity__dop-hyd_t0.5-5.0_rescale-False}}}
            \caption{\textbf{Cumulative number of detectable, periodically variable MBH Binaries in sensitivity versus redshift space}.  Doppler variable binaries (left) are considered independently from hydrodynamic ones (right).  The plotted values are all-sky detection rates for binaries with observer-frame periods between $0.5$ and $5.0 \, \yr$.  The sensitivities of CRTS, \mbox{$\fnusenscrts = 4\E{-28} \sfluxunits$} ($m_v \approx 19.5$), and LSST, \mbox{$\fnusenslsst = 3\E{-30} \sfluxunits$} ($m_v \approx 24.9$), are marked with dashed, grey lines.}
            \label{fig:6_num-var_sens_redshift}
        \end{figure*}

        The number of detectable, periodically variable MBH binaries are shown as a function of survey sensitivity and redshift in \figref{fig:6_num-var_sens_redshift}.  The detection of Doppler variable binaries (left panel) becomes plausible only for sensitivities $m_v \gtrsim 21$.  Hydrodynamic variables are much more common, and models predict that sources could be identifiable even near $m_v \approx 18.5$.  At low sensitivities, observable sources from both models of variability are likely to appear near $z \approx 0.1$.  With much higher sensitivities, sources could be identified out to $z \approx 1.0$, but likely not much farther.  To study the observability of variables in more detail, we focus on two sensitivities representative of CRTS and LSST, \mbox{$\fnusenscrts = 4\E{-28} \sfluxunits$} ($m_v \approx 19.5$) and \mbox{$\fnusenslsst = 3\E{-30} \sfluxunits$} ($m_v \approx 24.9$) respectively, which are marked in \figref{fig:6_num-var_sens_redshift} with dashed, grey lines.  At the sensitivities of CRTS, our models predict $0.5$ and $10$ Doppler and hydrodynamic variables respectively, to be observable on the full sky; for LSST those numbers increase substantially to $30$ and $200$.

        % Fig 7 - Doppler Number-Density vs. Redshift, by AGN, all-bins, flux-bins, var-bins
        \begin{figure}
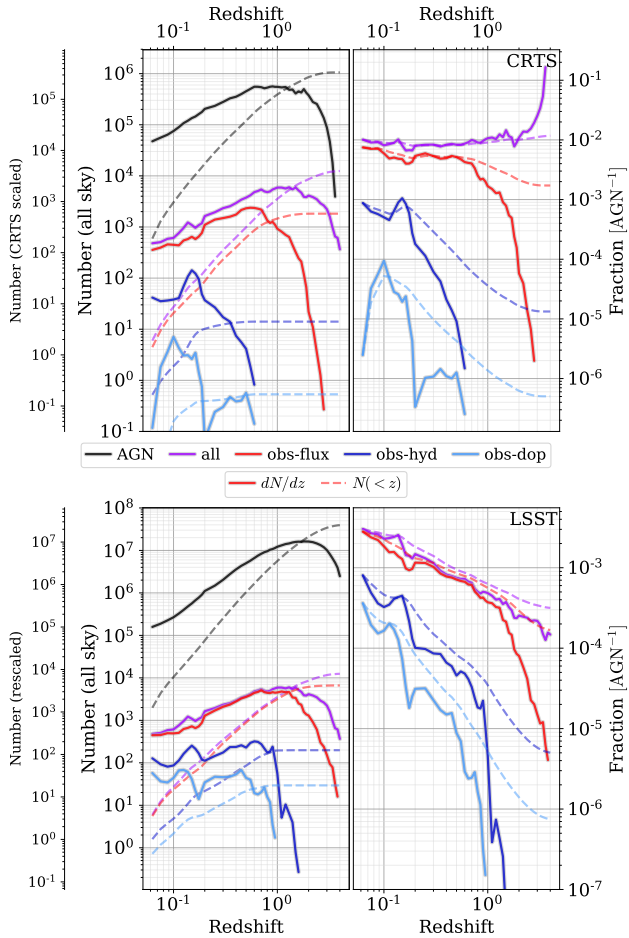

            \includegraphics[width=\columnwidth]{{{figs/7_num-variables_vs_redshift__dop-hyd_crts-lsst_t0.5-5.0}}}
            \caption{\textbf{Observability of MBH Binaries versus redshift} for CRTS (top) and LSST (bottom).  Observable AGN are shown in black, while all binaries are in purple, and those above the flux limit of each instrument are in red.  Binaries with observable variability are shown in light blue for Doppler variables, and dark blue for hydrodynamic variables.  The left panels give the absolute number of systems in the full sky, while the right panels give the the number of binaries as a fraction of observable AGN.  Differential distributions are shown with solid lines, and cumulative ones with dashed lines.  Values for binaries include all systems with observer-frame periods between $0.5$ and $5.0 \, \yr$.}
            \label{fig:7_num-var_redshift}
        \end{figure}

        % Introduce Number-vs-Redshift distributions; discuss all-binary and flux-obs rates
        We compare the populations of AGN, MBHBs, and observably variable binaries in \figref{fig:7_num-var_redshift}.  The black curves show the distribution of AGN, while all binaries are shown in purple, and those above the flux limit of each instrument are shown in red.  Binaries which are observably variable are plotted in light and dark blue for Doppler and hydrodynamic variability respectively. The most accurate predictions from our models are likely the number of binaries per observable AGN (right panels) which should reduce systematic uncertainties in the typical luminosities of our simulated AGN and binaries.  For convenience, we also include a second, left y-axis with the predicted number of sources rescaled to the number of studied, spectroscopically confirmed AGN in CRTS\footnote{In other words, the `CRTS scaled' y-axis is shifted such that the total number of predicted AGN matches the observed $3.3\E{5}$ of CRTS; see \refeq{eq:rescale}.}.

        % Redshift dependence of all and flux-observable binaries
        From \figref{fig:7_num-var_redshift}, we can see that the number of all binaries (purple) closely traces that of AGN, except two orders of magnitude fewer.  For $z \lesssim 0.6$, our models predict that most MBHBs are above the flux limit for CRTS (top, red) which implies that on the order of $1\%$ of AGN out to similar distances could be in binaries---$\approx 10^3$ systems.  The number of observably bright binaries falls off rapidly above $z \approx 1$ for CRTS, and above $z \approx 1.5$ for LSST.  In the full volume ($z \approx 4.0$), the fraction of AGN in binaries decreases to about $10^{-3}$ for CRTS, and $10^{-4}$ for LSST.

        % Overall rate of hydro-variables for CRTS and LSST
        Out to a redshift of $z \approx 0.2$, roughly $10\%$ of binaries observable above the CRTS flux limit are also identifiable as hydrodynamic variables (dark blue), corresponding to $\approx 10^{-3} \, \invagn$ hydrodynamically variable binaries.  This decreases to $\approx 10^{-4.5} \, \invagn$ at $z\approx 1$, and $\approx 10^{-5} \, \invagn$ in the full volume.  Scaling to the number of monitored CRTS AGN, our models predict $5$ hydrodynamically variable binaries should be visible.  For LSST, $\approx 25\%$ of binaries above the flux limit are also identifiable as hydrodynamically variable systems to $z \approx 0.2$, and $\approx 10\%$ to $z \approx 1$.  The total number of systems with hydrodynamic-variability above the LSST threshold is predicted to be $\approx 2\E{2}$ in the full sky, or $\approx 100$ assuming twice the completeness of CRTS (i.e.~complete for roughly half of the sky).

        % Overall rate of doppler-variables for CRTS and LSST
        The number of binaries with Doppler variability above the variability threshold (light blue) is drastically fewer than hydrodynamic ones, as only binaries with nearly-relativistic velocities produce sufficiently large brightness modulations.  Our models predict that, optimistically, of order unity Doppler-variable binaries should be present in the \citet{graham2015} CRTS data set, specifically an expectation value of $0.5$ all sky, and $0.2$ after rescaling to the number of CRTS AGN.  While Doppler variables are expected at a rate of $\sim 10^{-6} \, \invagn$ for both CRTS and LSST, this corresponds to a much more promising number of LSST sources: $30$ Doppler variable binaries all-sky, or $20$ after rescaling.

        % Fig 8 - Number-Density vs. Period, by total-masses
        \begin{figure}
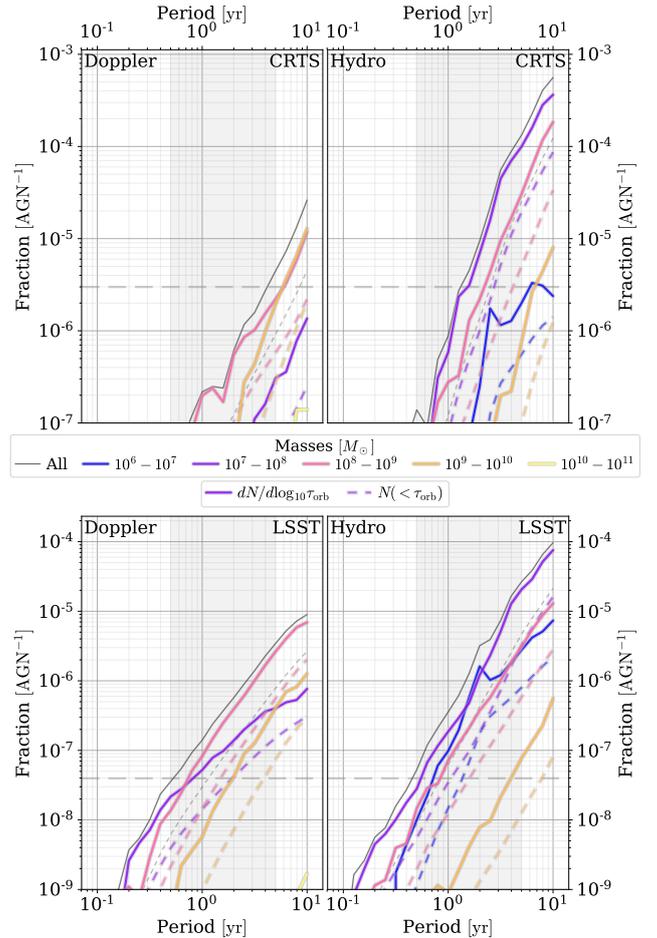

            \includegraphics[width=\columnwidth]{{{figs/8_num-variables_vs_period_for_mtot__dop-hyd_crts-lsst_per-agn-True}}}
            \caption{\textbf{Observability of Variable MBH Binaries versus period} for CRTS (top) and LSST (bottom).  Doppler variable systems are shown in the left column while hydrodynamic variables are shown on the right.  Binaries are broken into bins of different total masses indicated by the different colored lines, while the distribution of all binaries is in black.  Solid lines indicate distributions per log-decade of period, while dashed lines are cumulative.  The grey, horizontal dashed lines provide an estimate of the minimum observable occurrence rates, scaled to the CRTS completeness and twice that for LSST.}
            \label{fig:8_num-var_dop_period}
        \end{figure}

        % Introduce Distributions vs period, comment on period trends
        The predicted number of detectable variables are shown versus period in \figref{fig:8_num-var_dop_period}, grouped in bins of different total masses.  The horizontal dashed lines provide an estimate of the minimum plausibly-observable event rates.  For CRTS, this is the inverse of the number of monitored AGN from the variability survey, and for LSST we again rescale to twice the completeness of CRTS (see \refeq{eq:rescale}).  The shaded region highlights a decade of period between $0.5$ and $5.0 \, \yr$, corresponding to the values plotted \figref{fig:7_num-var_redshift} and representative of the identified CRTS candidates.  The number of binaries declines sharply with decreasing orbital period for all masses, reflecting the GW-hardening timescale, \mbox{$\tgw \propto \torb^{8/3}$}, by which nearly all binaries at these periods will be dominated  \citep[see, e.g.][]{hkm09, paper1}.

        % Period dependence and plausible periods of detection for CRTS and LSST
        The most obvious feature of \figref{fig:8_num-var_dop_period} is that binaries with both types of variability, and observed by either current CRTS-like instruments or even future LSST-like surveys, will be dominated by systems at the longest orbital periods.  This is unfortunate as the same is naturally expected from red-noise contaminated single AGN\footnote{Most AGN variability is well fit by a damped random walk, i.e.~red-noise at high frequencies, but flattening to white below some critical frequency.  While the transition is typically at inverse-months, the distribution has tails extending to many years containing binaries which would still seem red \citep{MacLeod2010}.}.  Our models suggest that Doppler-variable systems are only marginally observable by CRTS, with sources only likely to occur near and above $\per \approx 5 \, \yr$.  LSST on the other hand will likely detect systems down to $\per \approx 2 \, \yr$, where multiple orbital cycles could be observed, thereby decreasing the chance of red-noise contamination.  Hydrodynamic variables are far more common, and could be observed down to $\per \approx 1 \, \yr$ by CRTS and $\per \approx 0.5 \, \yr$ by LSST.

        % Mass dependence both Doppler and hydro, both CRTS and LSST
        Doppler variability depends on the orbital velocity, and thus favors larger total masses at a fixed orbital period.  This is clearly seen in \figref{fig:8_num-var_dop_period}, where systems detectable by CRTS are dominated by those in the $10^8 - 10^9 \, \msol$ and $10^9 - 10^{10} \, \msol$ mass bins.  Because LSST is much more sensitive, it begins to probe lower total masses, especially at smaller periods where the orbital velocity increases at fixed mass.  Apparent in the Doppler-variable LSST population (bottom left), lower mass systems are relatively more observable at shorter periods, where their velocities are larger.  Hydrodynamic variability has no explicit total-mass dependence, and thus those binaries tend to be dominated by the much more numerous, lower-mass systems.  Hydrodynamic variables are still not representative of all binary masses as the more massive systems can be more luminous, and thus observed in a larger volume.

    % 3.2
    \subsection{Observation Efficiency}
        \label{sec:res_dop_eff}

        % Fig 9 - Detection Efficiency: Binaries vs mtot and mrat, agn, flux and vars
        \begin{figure}
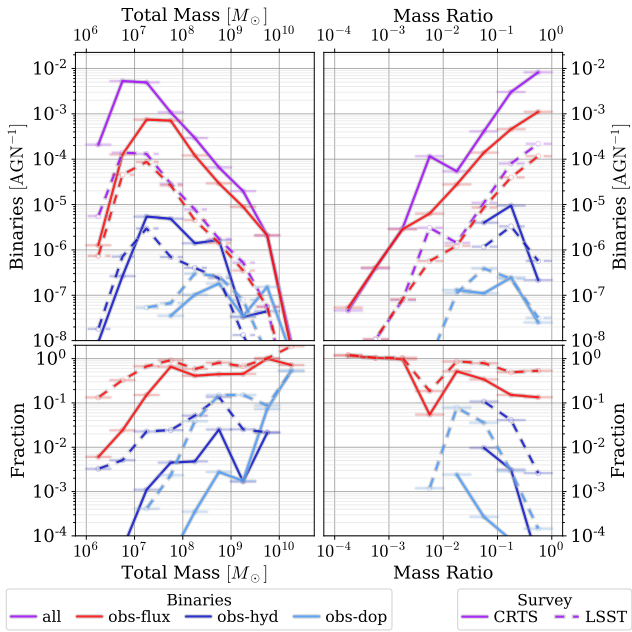

            \centering
            \includegraphics[width=\columnwidth]{{{figs/9_det-eff_mtot-mrat_dop-hyd_crts-lsst_t0.5-5.0}}}
            \caption{\textbf{Detection Efficiency of MBH Binaries as periodic variables} in bins of total mass (left) and mass ratio (right) for sensitivities of both CRTS (solid) and LSST (dashed).  The top panels show binary occurrence rates per observable AGN, and the bottom panels show the fraction of binaries which are observable: above each survey's flux-limit (red), and hydrodynamically variable (dark blue) and Doppler variable (light blue) above each survey's variability threshold.  Values are for binaries at all redshifts with observer-frame periods between $0.5$ and $5.0 \, \yr$.}
            \label{fig:9_det-eff}
        \end{figure}

        % Declining number vs increasing total-mass and decreasing mass-ratio
        To get a better sense of the fraction of binaries that are detectable, and their parametric dependencies, we plot detection efficiency versus total mass and mass ratio in \figref{fig:9_det-eff}.  The top panels, which show the number of binaries per AGN, demonstrate that the overall number of binaries falls quickly with increasing total mass (left), and thus also with decreasing mass ratio (right) as these quantities are strongly inversely-correlated.  Not only is the mass-function of MBH sharply decreasing with mass, but also for a fixed orbital period, higher mass systems harden faster---further decreasing their number\footnote{While this point does not effect our results, note again that some studies find that more massive binaries do not harden faster.}.  The lowest total-mass bin is an outlier, with noticeably fewer binaries due primarily to the mass cut at $10^6 \, \msol$, but also to the difficulty for low-mass systems to merge effectively and reach the binary separations corresponding to $\sim \yr$ periods.

        % Total-mass completeness fraction
        The lower panels of \figref{fig:9_det-eff} show the number of observable binaries (red and blues) divided by the total number of binaries in each bin.  This highlights that while observable binaries tend to follow the distribution of all binaries, their completeness drops significantly at lower masses.  Doppler-variable binaries (light blue) have an especially strong total-mass dependence, such that the efficiency of CRTS falls from $\approx 10^{-1}$ at $10^{10} \, \msol$ to well below $10^{-2}$ at $10^9 \, \msol$.  LSST maintains an efficiency of $\approx 10^{-1}$ down to roughly $10^{8.5} \, \msol$ before dropping.  While the peak recovery fraction of hydrodynamic variables (dark blue) is no better than for Doppler ones, the orders of magnitude higher efficiency for the most numerous binaries at $\lesssim 10^{8} \, \msol$ leads to vastly more, detectable hydrodynamically-variable binaries overall.

        % Hydrodynamic mass-ratio dependence
        Both variability mechanisms have strong and similarly shaped mass-ratio dependence, but largely for different reasons.  Our models assume that for mass ratios $q < \qmin$ (fiducially, $\qmin = 0.05$) there is no hydrodynamic variability as the secondary acts as a minor perturber \citep[see][]{farris201310}.  For mass ratios $q > \qmin$, however, there is no dependence in our model between the amplitude of variability and binary mass ratio, though a strong trend is evident in the detection efficiencies (dark blue, lower-right panel of \figref{fig:9_det-eff}).  While mass ratio does not affect the amplitude of variability, at high mass-ratios, $q > \qcrit$ (fiducially, $\qcrit = 0.3$), the period of variability is taken to be five times longer than the orbital period.  This means that high mass-ratio hydrodynamic variables observed at $\per \sim 5 \, \yr$, for example, are actually produced by the much smaller population of systems at $\per \sim 1 \, \yr$.  That shift in period leads to a significant drop in the number of hydrodynamic variables observed in the highest mass-ratio bin.  The tendency for higher mass-ratio systems to be lower in total mass likely also contributes to the high mass-ratio decline, as evidenced by a similar (though more subtle) decline for systems above the flux-limit (red).

        % Doppler mass-ratio dependence
        Binaries are easier to observe as Doppler variables when they have larger orbital velocities and thus smaller mass-ratios at a fixed total mass.  At the same time, at very low mass ratios, the variabilities which are produced by the secondary become washed out by the brighter primaries.  This leads to a strong peak in the detection efficiency of Doppler variables with mass ratio near $q \approx 0.03$ -- $0.07$.  For both Doppler and hydrodynamic variability, there is also a boost for systems near $q \approx 0.05$ where the accretion rate onto the secondary peaks (see \figref{fig:1_acc_ratio}).

    % 3.3
    \subsection{Model Effects and Parameters}
        \label{sec:app_pars}

        \newcommand{\modqaccrete}{`$q$-accrete'}

        % Fig 10 - Model Comparisons / Parametric dependencies
        \begin{figure}
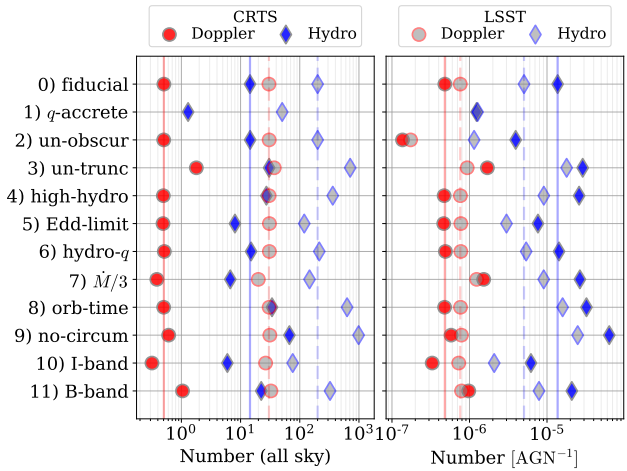

            \centering
            \includegraphics[width=\columnwidth]{{{figs/10_simple-comparisons}}}
            \caption{\textbf{Comparison of detection rates for parametric changes from our fiducial model} for CRTS (colored faces) and LSST (colored edges).  The left panel shows all-sky expected detection rates, and the right panel shown rates normalized to the number of observable AGN by each instrument.  Red points indicate Doppler variables while blue points indicate hydrodynamic ones.  The vertical lines denote the detection rates of the fiducial models.  See the surrounding text for a description of each alternative model.  The change in detection rates is typically well under an order of magnitude, except for the \modqaccrete{} model which eliminates all Doppler detections and decreases hydrodynamic ones by $\approx 10\times$ and $\approx 3 \times$ for CRTS and LSST respectively.}
            \label{fig:10_rate-comp}
        \end{figure}

        Our models of binary AGN variability include a variety of effects and uncertain parameters.  Figure~\ref{fig:10_rate-comp} compares our fiducial detection rates (top row and vertical lines) with variations from different parametric changes to our models.  Each row gives the expected all-sky (left) and per-AGN (right) detection rates for each alternative configuration.  Doppler variables are shown in red and hydrodynamic variables in blue.  The variations we consider are meant to illustrate both parametric uncertainties and the significance of particular physical effects.  Each alternative model is described below.

        \begin{itemize}
            \item \textbf{1) \modqaccrete{}}: the accretion rate of each component MBH is scaled proportionally to its mass, i.e.~\mbox{$\mdotrat \equiv \dot{M}_2 / \dot{M}_1 = q$}.  While the relative accretion rate has been studied extensively in the planetary and binary communities, a consensus has not been reached.  For both Doppler and hydrodynamic variability, the secondary MBH is almost always the source of observable periodicity, making the particular form of the accretion partition function (\figref{fig:1_acc_ratio}) important.  We test this, very simple model to explore our sensitivity to the accretion function.  Scaling the accretion rate of each MBH to its mass produces significantly lower secondary accretion rates than using the \citealt{farris201310} model.  Because Doppler variable sources are preferentially lower mass-ratio, the decreased accretion rate onto the secondary means that its variability is hidden by the brighter primary, and effectively never detectable.  The rate of hydrodynamic variable detections is decreased by an order of magnitude for CRTS and roughly a factor of three for LSST.

            \item \textbf{2) `un-obscur'}: neglecting AGN obscuration.  Interestingly, the overall number of detectable variables is virtually unaffected by the presence of obscuration, while, without obscuration, the total number of observable AGN is increased by a factor of a few.  This highlights the fact that the limiting condition for observing binaries photometrically tends to be in the variability sensitivity as opposed to the overall flux sensitivity.  Improving the minimum detectable variation amplitudes (i.e.~the floor variability sensitivity, $\dfnumin$) could significantly increase the number of binary detections, even at a fixed flux sensitivity.

            \item \textbf{3) `un-trunc'}: each MBH's accretion disk extends to indefinitely, instead of being truncated by the presence of the companion.  By definition this model only affects the observability of binaries and not single AGN.  All binary detection rates are increased by the presence of larger emitting regions, although Doppler-variable binaries detected by LSST are only slightly increased, implying that the intrinsic variability amplitude is the limiting factor and not the brightness of the secondary.  Doppler and hydrodynamic detections by CRTS are increased by about a factor of four and two respectively, while LSST hydrodynamic detections are increased by roughly a factor of three.  This model demonstrates the importance of disk truncation.  While AGN in the ADAF state (and thus, typically at low accretion rates) are unlikely to contribute significantly to the detectable population, better understanding if and how truncation occurs will be important in determining their relative observability.

            \item \textbf{4) `high-hydro'}: the amplitude of hydrodynamic variability is increased from $\hvaramp = 1.5$ to $2.0$.  In the simulations of \citet{farris201310}, the authors find different variability amplitudes for different mass-ratios, generally varying from $\approx 1.5$ to $\approx 3.0$.  In this model, we double the brightness increase that occurs from hydrodynamic variability and find that it produces only a moderate increase in hydrodynamic detections, $\approx 90\%$.  The systems which become detectable in this model are typically in either small mass-ratio binaries, or at lower periods, where truncation of each MBH's circum-single disk hampers its observability.

            \item \textbf{5) `Edd-limit'}: the accretion rate in each circum-single disk is limited to Eddington.  In our fiducial model, only the overall accretion rate to both MBHs is Eddington limited.  Because the secondary accretion rate is larger than the primary's for $q \sim 0.1$, it can exceed the Eddington limit individually.  Conceptually, the `Edd-limit' model assumes that the gas inflow is regulated effectively even at small scales, or alternatively that the radiative efficiency does not increase for Eddington fractions above unity\footnote{Some theoretical and numerical work have shown that the radiative efficiency increases only logarithmically for super-Eddington accretion rates in the `slim'-disk regime \citep[e.g.][]{1980AcA....30....1J, 2005gbha.conf..257A, 2009ApJS..183..171S}.}.  Estimated Doppler variable detections by CRTS and LSST are both decreased negligibly.  Hydrodynamic variable detection rates are decreased by $\approx 50\%$ for both instruments.  Hydrodynamically variable systems are more sensitive to this limit likely due to their tendency to be at slightly higher mass-ratios ($q \approx 0.1$), which are closer to the peak of the accretion partition function.

            \item \textbf{6) `hydro-$q$'}: The minimum mass-ratio for variability is set to $\qmin = 0.0$ (fiducial $\qmin = 0.05$).  Because the contribution from the secondary MBH in lower mass ratio systems is very small, they tend to be unobservable even if they produce hydrodynamic variability, and thus this model shows an effectively negligible change to detection rates.

            \item \textbf{7) `$\dot{M}/3$'}: The accretion rates from Illustris are uniformly scaled down by a factor of three.  The AGN luminosity function constructed from Illustris is generally consistent with observations, but noticeably over-predicts the number of observable systems.  The luminosity functions fit better if the accretion rates (or radiative efficiencies) are systematically decreased by $\approx 50\%$, but in that case the predicted number of AGN detectable by CRTS becomes too low.  This model attempts to quantify the uncertainty in the normalization of the Illustris luminosity function by systematically and significantly decreasing MBH accretion rates.

                For CRTS, the all sky predicted number of identifiably variable systems decreases from $0.51$ to $0.39$ (Doppler) and $14$ to $6.7$ (hydrodynamic).  At the same time, the predicted number of CRTS detectable AGN decreases from $1.1\E{6}$ to $2.5\E{5}$, which is inconsistently low compared to the number monitored by CRTS.  Rescaling to the actual number of CRTS observed AGN leads to an increase in the predicted variable detection rates: from $0.16$ to $0.51$ (Doppler) and $4.6$ to $8.9$ (hydrodynamic).  Systematically decreasing accretion rates leads to fewer observable variables in our simulated sky, and far fewer AGN.  Assuming that the number of observable variables \textit{per AGN} is the most accurate predictor, however, actually leads to an increase in predicted detection rate for CRTS and LSST from the low accretion-rate model.

            \item \textbf{8) `orb-time'}: hydrodynamic variability always occurs at the orbital period, instead of shifted to longer periods for $q > \qcrit = 0.3$.  Some hydrodynamic simulations suggest that high mass-ratio systems will exhibit variability at periods roughly five times longer than the binary period, corresponding to the orbital time of an over-density of material at the outer edge of the gap in the disk.  For these systems, observed periodicity at $\sim 1 \, \yr$ is actually coming from sources with orbital periods of $\sim 0.2 \, \yr$, which are far less numerous.  If variability always occurs at the orbital period, the number of hydrodynamically variable detections double for both CRTS and LSST.  Specifically, rescaled detection rates increase from $4.6$ to $11$ for CRTS, and from $130$ to $400$ for LSST.  It's worth noting that in the \citet{farris201310} simulations, while variations are \textit{predominantly} at $\approx 5 \times$ the orbital period, there is still a component directly at the orbital period itself which could be identifiable for sufficiently high signal-to-noise systems.  To be conservative, we use the time-shifted configuration as our fiducial model.

            \item \textbf{9) `no-circum'}: emission from the circumbinary portion of the accretion disk is neglected.  Without the circumbinary disk, fractional brightness variations increase as there is less steady emission to compete with.  This increases Doppler detection rates by only $\approx 20\%$ for CRTS and negligibly for LSST, but increased Hydrodynamic detection rates by almost a factor of four for both CRTS and LSST.  Circumbinary emission seems to primarily hamper the detection of near equal-mass binary systems, which are significantly less important for Doppler detection rates.  This model highlights that circumbinary emission is an important source of flux to consider.  Additionally, recent simulations suggest that the circumbinary portion of the disk can also contribute to periodic variability signatures \citep{2018MNRAS.476.2249T} which is not considered here.

            \item \textbf{10) `B-Band'} and \textbf{11) `I-Band'}: spectra are sampled in the rest-frame B-Band and I-Band instead of the V-Band.  For both CRTS and LSST, more variables are detectable in the B-Band and fewer in the I-Band.  Doppler detections by CRTS increase by almost a factor of two, and negligibly for LSST, while hydrodynamic detections increase by $\approx 40\%$ for both instruments.  Keeping in mind that our models do not take into account factors such as color-dependent extinction, this effect seems to be driven by the AGN circum-single disks being intrinsically brighter on the bluer side of the spectrum.

        \end{itemize}

% Sec 4
\section{Discussion}
    \label{sec:disc}

    In this paper, we make predictions for the electromagnetic detection of MBH binaries as photometrically periodic AGN.  We use a population of binaries drawn from cosmological hydrodynamic simulations of MBHs and galaxies, evolved using semi-analytic, post-processing models of the detailed merger physics.  Employing synthetic AGN spectra, along with models of both Doppler and hydrodynamic variability, we have calculated detection rates for the flux and variability sensitivities of CRTS and LSST.  Here we present the results and implications of our study, after first discussing their limitations.

    % Sec 4.1
    \subsection{Caveats}
        \label{sec:disc_cavs}

        Numerous limitations exist in our current methods, both in terms of our binary populations and our models of variability.  In the former class, while the masses of MBHs in Illustris nicely reproduce the observed, redshift zero BH--galaxy scaling relations \citep{sijacki2015}, there is still significant uncertainty in the full distribution of MBH masses \citep[e.g.][]{mcconnell2013}.  The MBH accretion rates have been calibrated to produce accurate masses and reproduce observational, bolometric luminosity functions \citep{sijacki2015}.  Using spectral models, and a simple model of obscuration, applied to the entire population of Illustris MBHs, we predict a total number of observable AGN (see \tabref{tab:rates}) which are consistent with CRTS observations, but the AGN luminosity function from our model over predicts the observed relation from \citet[][reproduced in \figref{fig:4_agn_lumfunc}]{Hopkins2007}.  While we overestimate the luminosity function even for the brightest systems, we still suffer from small number statistics and incompleteness in the most massive MBHs, due to the finite volume of the Illustris box.

        Instead of using bolometric luminosities and corrections, or characteristic spectral indices, we have constructed full spectral models for each of the MBHs in our binary populations.  Still, these spectra are highly simplified in the complex and actively developing field of AGN emission.  Perhaps the most important deficiency of our spectral models are the lack of any lines, color-dependent extinction, or non-thermal contributions from outside of the disk.  We also do not consider any intrinsic AGN variability.  Full spectroscopic observations of AGN, including variations between observing epoch, should be incorporated into calculations to carefully consider the effectiveness with which periodic photometric variability can be accurately classified.

        We have also relied very strongly on the results of \citet{farris201310}, which use 2D, isolated, purely-hydrodynamic simulations.  Other groups \citep[e.g.][]{Cuadra200809, Roedig201202, 2012ApJ...749..118S} have supported the \citet{farris201310} results whose conclusions seem robust for their simulated conditions.  Accounting for thick-disk accretion (and mass flow out of the disk plane, possibly enhanced by magnetic fields and radiation), and turbulent flows with varying inflow rates on large scales, are likely to affect detailed predictions.  Using the \citet{farris201310} accretion rates also produces an inconsistency in our models: the large accretion rates to $q \sim 0.1$ binaries, combined with the typically large systemic accretion rates in Illustris, imply that binaries will grow towards $q \sim 1$ relatively quickly.  Naively, the e-folding time for the secondary at these mass ratios is often as short as $10$ Myr.  At the same time, the accretion is typically super-Eddington in these systems, and whether those accretion rates accurately correspond to mass-growth rates (i.e.~neglecting outflows) is unclear.  If the secondary MBHs were allowed to grow self consistently, it would likely decrease our predictions for both types of variables.  At the same time, some studies in the context of planetary systems \citep[e.g.][]{1999ApJ...526.1001L} find that even at relatively extreme mass-ratios (e.g.~$q \sim 10^{-3}$),  the accretion rate onto the secondary can still be a significant fraction of the total accretion rate.  If accretion rates remain high at low mass-ratios, it could significant increase the incidence of detectable systems.

        We expect many of the simplifications and uncertainties in our models to tend towards \textit{fewer} systems being observable as variables.  For example, thick-disk and turbulent accretion flows are more likely to smooth out the periodic variations in emission rather than enhance them.  Light from AGN host galaxies will also produce an additional background from which variability must be disentangled.  AGN are also known to exhibit strong intrinsic variability, especially at long periods which easily mimics periodicity.  While our model for variability sensitivity is based on observational studies of signal identification, it is an extremely simplistic accounting of a very difficult task, as shown in the detailed analyses of \citet{graham2015}, \citet{charisi2016}, and \citet{Liu201609} which include careful treatments of noise.

    % Sec 4.2
    \subsection{Conclusions}
        \label{sec:disc_conc}

        A summary of expected detection rates for variability periods from $0.5$ -- $5$ yr are presented in \tabref{tab:rates}.  Our models predict that MBH binaries should be detectable at rates of roughly $5\E{-7} \, \invagn$ and $10^{-5} \, \invagn$ for Doppler and hydrodynamic variability, respectively.  In our simulations, this corresponds to all-sky rates of $0.5$ and $10$ sources at the CRTS sensitivity, while scaling to the number of CRTS-monitored AGN yields $0.2$ and $5$ binaries.  For the expected sensitivity of LSST, and assuming twice the completeness of AGN confirmation, we predict $20$ and $100$ Doppler and hydrodynamic binaries to be observable.  Additional data is presented in \secref{sec:app} and online to facilitate detection-rate predictions for optical surveys with different sensitivities and durations.

        \renewcommand{\arraystretch}{2.0}
        \setlength{\tabcolsep}{4pt}
        \begin{table*}
            \begin{center}

            \begin{tabular}{ r | l l | l l | l l }
                                & \multicolumn{4}{c|}{Observable, Variable Binaries}                                                                                                 & \multicolumn{2}{c}{AGN}                                           \\
                                & \multicolumn{2}{c|}{Doppler}                                             & \multicolumn{2}{c|}{Hydrodynamic}                                       &                                 &                                 \\
            Number              & CRTS                               & LSST                                & CRTS                               & LSST                               & CRTS                            & LSST                            \\ \hline
            All Sky		        & $5\E{-1}$ $\left(_{3\E{-1}}^{2\E{+0}}\right)$ & $3\E{+1}$  $\left(_{2\E{+1}}^{4\E{+1}}\right)$ & $1\E{+1}$ $\left(_{6\E{+0}}^{7\E{+1}}\right)$ & $2\E{+2}$ $\left(_{8\E{+1}}^{1\E{+3}}\right)$ & $1\E{6}$ $\left(_{3\E{5}}^{4\E{6}}\right)$ & $4\E{7}$ $\left(_{2\E{7}}^{2\E{8}}\right)$ \\
            $\textrm{AGN}^{-1}$ & $5\E{-7}$ $\left(_{1\E{-7}}^{2\E{-6}}\right)$ & $8\E{-7}$  $\left(_{2\E{-7}}^{1\E{-6}}\right)$ & $1\E{-5}$ $\left(_{4\E{-6}}^{6\E{-5}}\right)$ & $5\E{-6}$ $\left(_{1\E{-6}}^{2\E{-5}}\right)$ & $1\E{0}$ $\left(-\right)$                  & $1\E{0}$ $\left(-\right)$                  \\
            Scaled		        & $2\E{-1}$ $\left(_{5\E{-2}}^{6\E{-1}}\right)$ & $2\E{+1}$  $\left(_{6\E{+0}}^{5\E{+1}}\right)$ & $5\E{+0}$ $\left(_{1\E{+0}}^{2\E{+1}}\right)$ & $1\E{+2}$ $\left(_{4\E{+1}}^{6\E{+2}}\right)$ & $3\E{5}$ $\left(-\right)$                  & $3\E{7}$ $\left(_{3\E{7}}^{4\E{7}}\right)$ \\

            \end{tabular}

            \caption{\textbf{Expected observability of MBH binary, periodically variable AGN}.
                The first four columns give the expected detection rates of Doppler and Hydrodynamic variable binaries with periods between $0.5$ and $5.0\,\yr$. The last two columns give the expected detection rates for (single) AGN.  Each cell includes in parenthesis the range of values from the models discussed in \secref{sec:app_pars}.  The first row is the all-sky prediction from our simulations, while the second row is normalized to the predicted number of detectable AGN for each instrument.  The last row rescales our results based on the number of AGN monitored in the CRTS variability study \citep[$\approx 3.3\E{5}$;][]{graham2015}, and assuming twice that completeness for LSST.}
            \label{tab:rates}
            \end{center}
        \end{table*}

        Our predictions for current instruments are significantly lower than the rate of candidates put forward by \citet{graham2015} and \citet{charisi2016}.  Our models suggest that having multiple Doppler variables are unlikely in CRTS, and only a small fraction of the candidates can be explained as hydrodynamic variables.  This is consistent with gravitational wave limits which imply that the published candidates contain false positives \citep{sesana201703}.  But our predictions do indicate that there should exist numerous true MBH binaries within the candidate populations.  The systems painstakingly identified by \citet{graham2015} and \citet{charisi2016} provide an extremely valuable opportunity to identify examples of MBHBs.  The candidates deserve significant followup to find the binaries they contain, as no examples of gravitationally-bound MBHBs have been confirmed to date.  Additionally, the candidates put forward are very convincing in demonstrating periodic variability \textit{above that produced by the best fitting models of intrinsic AGN variability}.  The characterization and study of the mechanisms producing false positives thus presents an interesting opportunity to not only better identify binaries, but also to explore the fundamental accretion processes at play.

        Simply extending the temporal baselines of candidate observations will provide a determinant in distinguishing red-noise contaminated systems.  Candidates exhibiting red-noise fluctuations misconstrued as periodicity are expected to eventually deviate from sinusoidal behavior.  Unfortunately, this test is not without issue.  Disk turbulence and time-varying feeding rates of gas can not only introduce their own luminosity fluctuations, but also decrease the coherent, periodic variations from a binary.  Some simulations also see accretion alternate from primarily feeding one MBH to then predominantly feeding the other \citep[e.g.][]{2015MNRAS.448.3545D, 2018ApJ...853L..17B}, even in otherwise smooth disks.  These factors introduce a significant complication in separating AGN with significant red-noise from those which are binaries, but exhibit excursions from periodicity.  While we find Doppler variables to be intrinsically very rare, the characteristic spectral dependence of their variations is a valuable test of their origin \citep{D'Orazio201509}.  Systems which may be seen edge-on, like Doppler variables, can also produce periodic lensing spikes \citep{2017PhRvD..96b3004H, 2018MNRAS.474.2975D}.  While our analysis does not consider identification of systems simultaneously exhibiting both Doppler and hydrodynamic variability, these phenomena could be coincident and observationally distinguishable.  It is also worth considering `triggered' searches, for example: from a candidate Doppler-variable binary, a survey could search for signs of hydrodynamic variability at multiples of the orbital period with boosted signal-to-noise.  Ultimately, we expect that time-varying spectroscopic features of binarity (e.g., \citealt{Comerford200810}; but see also \citealt{Eracleous1997}) may be the most robust identifier of spatially unresolved MBHBs.

        Our results offer hints at what candidate system parameters may be most indicative of true binarity.  We find that neither Doppler nor hydrodynamic variables are likely to be observed much beyond $z \approx 1$, as variability amplitudes are simply too low to be distinguished at larger redshifts.  Out to a redshift $0.6$, our models suggest that almost $1\%$ of AGN could harbor binaries, although only a small fraction are identifiable as such.  To plausibly detect hydrodynamically variable binaries, a survey must have a sensitivity of $m_v \gtrsim 18.5$, and $m_v \gtrsim 21$ for Doppler variable binaries.  Owing to the strong period dependence of the GW hardening rate, binaries are most likely to be observed at longer periods---unfortunately the same trend as for red-noise contamination.

        The Doppler variability model depends on nearly relativistic velocities, which means that more massive binaries are strongly favored.  We expect both current and future instruments to detect Doppler variables predominantly above $10^8 \, \msol$, and LSST will likely see mostly $10^8$ -- $10^9 \, \msol$ binaries.  Hydrodynamic variability is insensitive to the total binary mass, and thus is dominated by the far more numerous systems at lower masses.  Below $\approx 10^7 \, \msol$ however, the lower luminosities begin to limit the sensitive volume.  For both CRTS and LSST, hydrodynamically variable binaries should be mostly between $\approx 10^7$ and $10^8 \, \msol$, although systems between $10^{6.5}$ and $10^9 \, \msol$ should be observable, and higher mass systems are likely too rare.

        Both Doppler and hydrodynamic variability strongly favor systems with mass-ratios $q \sim 0.1$, in our fiducial model.  In the case of Doppler variability, this trend is due to the larger orbital velocities of secondary MBHs with lower mass-ratios.  The peak mass-ratio sensitivity occurs near $q \approx 0.05$, as secondaries become too faint to produce observable variability when their masses become much lower.  The bias towards $q \approx 0.1$ is enhanced by the heightened accretion rates for secondary MBHs near that mass ratio seen by \citet{farris201310}.  Hydrodynamically variable binaries are most apparent near $q \approx 0.1$, both because of the enhanced accretion rate, and because systems with $q \sim 1$ have variability periods shifted largely to longer periods.  In binaries with $q \ll 1$, the secondary again becomes quite faint, and the variability induced in the accretion flow also begins to be negligibly small.

        Selection biases in binary parameters identifiable as variables is important in estimating the GWB implied by a given binary population.  \citet{sesana201703} show that the CRTS candidates are in tension with current GWB upper limits from PTAs.  The authors show that the tension can be decreased if the mass-ratio distribution of MBHBs is biased towards lower values, as systems with low $q$ produce weaker GW signals.  Our full binary populations have relatively flat mass-ratio distributions in log-space, but the selection effects for detecting variability prefer binaries with lower mass-ratios.  This observational bias has the opposite effect of an intrinsically low-$q$ population.  The high-$q$ systems, which dominate the GWB, are absent from the observational candidates, but contribute more significantly to the GWB.  The amplitude of this effect can be estimated as follows.  For a given population of candidates, their most likely binary parameters can be used to estimate a GW strain that they produce directly, $h_c^{direct}$.  Our results suggest that over the full range of masses, less than $10\%$ of binaries are observable.  Furthermore, these systems should have $q \lesssim 0.1$, while binaries with $q \approx 1$ are intrinsically at least three times more common.  Those nearer equal mass binaries produce GWs with strains roughly ten times larger\footnote{Strain is related to chirp-mass as, $h \propto \mchirp^{5/3}$, and the chirp-mass is $\mchirp \equiv M q^{3/5} / (1 + q)^{6/5}$.}.  All together, because GW strains add in quadrature, this implies that the true GWB amplitude is $h_c \approx 300^{1/2} h_c^{direct}$.

        In conclusion, our models are able to explain only a fraction of previously identified, periodically variable binary candidates.  The distributions of detectable-binary parameters that we find suggest that existing PTA constraints on the GWB also require a large fraction of candidates to be false positives.  On the other hand, our results suggest that many of the candidates in CRTS and PTF could indeed be true MBH binaries.  We have presented the parameters of variables which we expect to be MBHBs in the hope that the candidates can be followed up with additional photometric and spectroscopic monitoring to find the binaries they contain. Confirmed examples of year-period MBHBs would present a boon to AGN and MBH binary astrophysics.  Such systems contain key information on MBH growth, MBH binary evolution, and offer stringent constraints and insights into our predictions for low-frequency GW signals soon to be detectable by PTAs, and eventually by LISA.

% Acknowledgments
% ---------------
\section*{Acknowledgments}
    The authors are grateful for insightful conversations and comments from Laura Blecha and Daniel D'Orazio, and helpful feedback from Edo Berger, Daniel Eisenstein, and Avi Loeb.

    This research made use of \astropy, a community-developed core Python package for Astronomy \citep{astropy2013}, in addition to \scipy~\citep{scipy}, \ipython~\citep{ipython}, \& \numpy~\citep{numpy2011}.  All figures were generated using \matplotlib~\citep{matplotlib2007}.

    ZH was supported in part by NASA through grants 16-SWIFT16-0015 and NNX17AL82G and by NSF through grant 1715661.  AS is supported by the Royal Society.

% ---- Bibliography ----
\let\oldUrl\url
\renewcommand{\url}[1]{\href{#1}{Link}}

\quad{}
\bibliographystyle{mnras}
% \bibliography{refs}
\bibliography{references}

\onecolumn
\clearpage

\appendix

    % \section{Additional Material}
    \section{}
    \label{sec:app}

    To facilitate comparisons with surveys beyond CRTS and LSST, we present our predicted observations of MBH binaries as periodically-variable AGN in sensitivity--variability-timescale parameter space.  Figure~\ref{fig:a1} gives detection rates in units of $\invagn$, \figref{fig:a2} as a cumulative number of all-sky observable systems.  The associated tabular data for Figs~\ref{fig:a1}~\&~\ref{fig:a2} are included online.  To estimate the number of detectable systems for a particular survey, the limiting variability period that it is sensitive to must be determined.  Nyquist sampling implies the maximum period is at most the total temporal baseline of the survey, but likely numerous complete cycles of variability should be observed to reject red-noise contamination.  Multiplying the predicted rates in \figref{fig:a1} by the actual number of AGN observed by a particular survey, as opposed to using \figref{fig:a2} directly, will provide a more robust prediction as it decreases systematic uncertainties from the simulated AGN luminosity function.  \vspace{0.1in}

    {
    \centering
            \includegraphics[width=0.95\columnwidth]{{{figs/num-variables_vs_sensitivity-period__dop-hyd_per-agn-True}}}
            \captionof{figure}{\textbf{Predicted rate of observable binaries in sensitivity--period parameter space.}  Occurrence rates are given in units of $\invagn$, which leads to non-monotonic behavior as a function of sensitivity.  Cumulative binary occurrence rates are shown for variations between $0.1 \, \yr$ and the indicated periods.  The observational duration of a given survey determines the maximum variability period that it is sensitive to.}
            \label{fig:a1}

            \includegraphics[width=0.95\columnwidth]{{{figs/num-variables_vs_sensitivity-period__dop-hyd_per-agn-False}}}
            \captionof{figure}{\textbf{Predicted number of all-sky observable binaries in sensitivity--period parameter space.}}
            \label{fig:a2}
    }

    \twocolumn

    \label{lastpage}

\end{document}

%% file: CustComs.tex
\newcommand{\tr}[1]{\textrm{#1}}
\newcommand{\trt}[1]{\textrm{\tiny{#1}}}

\newcommand{\msol}{\tr{M}_{\odot}}

\newcommand{\bigt}{\ensuremath{\uptau}}

\newcommand{\E}[1]{\times\nobreak10^{#1}}

\newcommand{\logten}{\log_{10}}
\newcommand{\lr}[1]{\left(#1\right)}
\newcommand{\scale}[3][]{\lr{ \frac{#2}{#3} }^{#1}}

\newcommand{\mdot}{\dot{M}}
\newcommand{\mdotedd}{\dot{M}_\trt{Edd}}

\newcommand{\ledd}{L_\trt{Edd}}    % Eddington luminosity
\newcommand{\radeff}{\varepsilon_\trt{rad}}

\newcommand{\secref}[1]{\textsection\ref{#1}}
\newcommand{\figref}[1]{Fig.~\ref{#1}}
\newcommand{\refeq}[1]{{Eq.~\ref{#1}}}
\newcommand{\tabref}[1]{{Table~\ref{#1}}}

\newcommand{\rs}{R_\trt{s}}    % Stellar radius
\newcommand{\torb}{\bigt_\trt{orb}}

\newcommand{\ayr}{A_{\trt{yr}^{-1}}}        % GWB amplitude normalization at 1/yr
\newcommand{\astropy}{\texttt{Astropy}}
\newcommand{\matplotlib}{\texttt{matplotlib}}
\newcommand{\numpy}{\texttt{NumPy}}
\newcommand{\scipy}{\texttt{SciPy}}
\newcommand{\ipython}{\texttt{ipython}}

\newcommand{\fedd}{f_\trt{Edd}}    % Mdot-eddington factor for limiting

\newcommand{\mchirp}{\mathcal{M}}     % Chirp-mass
\newcommand{\volill}{V_\trt{ill}}   % Illustris volume

\newcommand{\thard}{\bigt_\tr{h}}

\newcommand{\thardij}{\bigt_{\tr{h},ij}}

\newcommand{\tgw}{\bigt_\tr{gw}}

\newcommand{\yr}{\,\mathrm{yr}}        % yr in math-mode
\newcommand{\eccinit}{e_0}

\newcommand{\tvar}{\bigt_\trt{var}}
\newcommand{\zmax}{z_\trt{max}}
\newcommand{\qcrit}{q_\trt{crit}}
\newcommand{\qmin}{q_\trt{min}}
\newcommand{\dl}{d_\trt{L}}   % luminosity distance
\newcommand{\mdotrat}{\lambda}

\newcommand{\sfluxunits}{\textrm{ erg/s/Hz/cm}^2}

\newcommand{\invagn}{\textrm{AGN}^{-1}}

\newcommand{\fracobs}{f_\trt{obs}}
\newcommand{\fracobsi}{f_{\trt{obs},i}}
\newcommand{\feddsys}{f_\trt{Edd,sys}}

\newcommand{\feddtwo}{f_{\trt{Edd},2}}

\defcitealias{hkm09}{HKM09}
\defcitealias{Begelman1980}{BBR80}
\defcitealias{Rosado1503}{RSG15}
\defcitealias{paper1}{Paper-1}
\defcitealias{paper2}{Paper-2}
\defcitealias{paper3}{Paper-3}
\defcitealias{paper4}{Paper-4}

%% file: JournalAbbrevs.tex
\def\aj{Astron. J.}              		   		% Astronomical Journal
\def\apj{Astrophys. J.}       		        	 	% Astrophysical Journal
\def\apjl{Astrophys. J. Lett.}             		% Astrophysical Journal, Letters
\def\pasj{PASJ}

\def\apjs{Astrophys. J., Suppl. Ser.}            % Astrophysical Journal, Supplement
\def\mnras{Mon. Not. R. Astron. Soc.}        % Monthly Notices of the RAS
\def\prd{Phys. Rev. D}      				% Physical Review D
\def\araa{Annu. Rev. Astron. Astrophys.}  % Annual Review of Astron and Astrophys
\def\nat{Nature}              				% Nature
\def\aap{Astron. Astrophys.}               		% Astronomy and Astrophysics

\def\memsai{Mem. Societa Astronomica Italiana}